\documentstyle[a4p,12pt,epsfig]{article}

\begin{document}
\newcommand {\epem}     {e$^+$e$^-$}
\newcommand {\qq} {q\overline{q}}
\newcommand {\gincl} {$g_{incl.}$}
\newcommand {\ecm} { E_{c.m.} }
\newcommand {\evis} { E_{vis.} }
\newcommand {\nqqg} { {\mathrm{N}}_{q\overline{q}g} }
\newcommand {\nqqgch} { {\mathrm{N}}_{q\overline{q}g}^{ch.} }
\newcommand {\nqq} { {\mathrm{N}}_{q\overline{q}} }
\newcommand {\ngg} { {\mathrm{N}}_{gg} }
\newcommand {\ng} { {\mathrm{N}}_{g} }
\newcommand {\nq} { {\mathrm{N}}_{q} }
\newcommand {\nqqch} { {\mathrm{N}}_{q\overline{q}}^{ch.} }
\newcommand {\nggch} { {\mathrm{N}}_{gg}^{ch.} }
\newcommand {\nchee} { {\mathrm N}^{\,ch.}_{ 
           {\mathrm e}^+{\mathrm e}^- } }
\newcommand {\lqq} { {L}_{q\overline{q}} }
\newcommand {\ktle} { k_{\perp,Le} }
\newcommand {\ktlu} { k_{\perp,Lu} }
\newcommand {\ktgluon} { k_{\perp,gluon} }
\newcommand {\ptle} { p_{\perp,Le} }
\newcommand {\ptlu} { p_{\perp,Lu} }
\newcommand {\kperp} { k_{\perp} }
\newcommand {\ycut} { y_{cut} }
\newcommand {\dsign} { d_{sign} }
\newcommand {\caa} { C_{\mathrm A} }
\newcommand {\cff} { C_{\mathrm F} }
\newcommand {\nff} { n_f }
\newcommand {\sqq} { s_{q\overline{q}} }

\begin{titlepage}
\noindent
\begin{center}  {\large EUROPEAN ORGANIZATION FOR NUCLEAR RESEARCH }
\end{center}

\bigskip\bigskip\bigskip
\begin{tabbing}
\` CERN-EP-2001-076 \\
\` 25 October 2001 \\
\end{tabbing}

\bigskip\bigskip

\begin{center}{\LARGE\bf
Particle multiplicity of unbiased gluon jets
from {\epem} three-jet events
}
\end{center}

\begin{center}
{\Large
The OPAL Collaboration
}
\end{center}

\begin{center}{\large\bf  Abstract}\end{center}
\bigskip
\noindent
The charged particle multiplicities of two- and
three-jet events from the reaction
{\epem}$\,\rightarrow\,$Z$^0$$\,\rightarrow\,$$hadrons$
are measured for Z$^0$ decays to light quark (uds) flavors.
Using recent theoretical expressions to account for
biases from event selection,
results corresponding to unbiased gluon jets are extracted
over a range of jet energies from about 11 to 30~GeV.
We find consistency between these results and direct measurements
of unbiased gluon jet multiplicity from $\Upsilon$ 
and Z$^0$ decays.
The unbiased gluon jet data
including the direct measurements
are compared to corresponding results for quark jets.
We perform fits based on analytic expressions for particle
multiplicity in jets to determine
the ratio $r$$\,\equiv\,$$\ng/\nq$
of multiplicities between
gluon and quark jets as a function of energy.
We also determine the ratio of slopes,
$r^{(1)}$$\,\equiv\,$$({\mathrm{d}}\ng /{\mathrm{d}}y)
/({\mathrm{d}}\nq / {\mathrm{d}}y)$,
and of curvatures,
$r^{(2)}$$\,\equiv\,$$({\mathrm{d}}^2\ng /{\mathrm{d}}y^2)
/({\mathrm{d}}^2\nq / {\mathrm{d}}y^2)$,
where $y$ specifies the energy scale.
At 30~GeV,
%a typical scale for the gluon jets in our study, 
we find $r$$\,=\,$$1.422\pm0.051$,
$r^{(1)}$$\,=\,$$1.761\pm0.071$ and
$r^{(2)}$$\,=\,$$1.98\pm0.13$,
where the uncertainties are the statistical and systematic
terms added in quadrature.
These results are in general agreement with theoretical
predictions.
In addition,
we use the measurements of the energy dependence of
$\ng$ and $\nq$ to determine an effective value
of the ratio of QCD color factors,
$\caa/\cff$.
Our result,
$\caa/\cff$$\,=\,$$2.23\pm0.14\,$(total),
is consistent with the QCD value of 2.25.

\vspace*{3cm}
\begin{center}{\large
(Submitted to Eur. Phys. J. C)
}\end{center}

\end{titlepage}

\begin{center}{\Large        The OPAL Collaboration
}\end{center}
{\small
\begin{center}{
%begin authorlist PLEASE DO NOT DELETE THIS COMMENT
G.\thinspace Abbiendi$^{  2}$,
C.\thinspace Ainsley$^{  5}$,
P.F.\thinspace {\AA}kesson$^{  3}$,
G.\thinspace Alexander$^{ 22}$,
J.\thinspace Allison$^{ 16}$,
G.\thinspace Anagnostou$^{  1}$,
K.J.\thinspace Anderson$^{  9}$,
S.\thinspace Arcelli$^{ 17}$,
S.\thinspace Asai$^{ 23}$,
D.\thinspace Axen$^{ 27}$
% ...
%end authorlist PLEASE DO NOT DELETE THIS COMMENT
}\end{center}
%begin institutes
$^{  1}$School of Physics and Astronomy, University of Birmingham,
Birmingham B15 2TT, UK
\newline
$^{  2}$Dipartimento di Fisica dell' Universit\`a di Bologna and INFN,
I-40126 Bologna, Italy
\newline
$^{  3}$Physikalisches Institut, Universit\"at Bonn,
D-53115 Bonn, Germany
\newline
% ...
$^{  5}$Cavendish Laboratory, Cambridge CB3 0HE, UK
\newline
% ...
$^{  9}$Enrico Fermi Institute and Department of Physics,
University of Chicago, Chicago IL 60637, USA
\newline
% ...
$^{ 16}$Department of Physics, Schuster Laboratory, The University,
Manchester M13 9PL, UK
\newline
$^{ 17}$Department of Physics, University of Maryland,
College Park, MD 20742, USA
\newline
% ...
$^{ 22}$Department of Physics and Astronomy, Tel Aviv University,
Tel Aviv 69978, Israel
\newline
$^{ 23}$International Centre for Elementary Particle Physics and
Department of Physics, University of Tokyo, Tokyo 113-0033, and
Kobe University, Kobe 657-8501, Japan
\newline
% ...
$^{ 27}$University of British Columbia, Department of Physics,
Vancouver BC V6T 1Z1, Canada
\newline
% ...
$^{  0}$ now at University of Kansas, Dept of Physics and Astronomy,
Lawrence, KS 66045, USA
\newline
% ...
%end notes
}

\clearpage\newpage

\section{Introduction}

The mean charged particle multiplicity of a gluon jet
has often been measured in the annihilation of an 
electron and positron to hadrons,
{\epem}$\,\rightarrow\,$$hadrons$.
The usual method
(see for example~\cite{bib-hrs85}-\cite{bib-delphi96})
is to select three-jet quark-antiquark-gluon $q\overline{q}g$ 
final states for which the events and individual jets 
are defined using a jet algorithm such as the
Durham~\cite{bib-durhamjf} or Luclus~\cite{bib-luclusjf}
jet finder.
The particle multiplicity of a jet 
%(henceforth, the ``jet multiplicity'')
determined with this technique
is found to depend on which algorithm is employed.
Therefore, these jets and the associated $q\overline{q}g$ 
events are called ``biased.''

In contrast,
theoretical calculations usually define gluon
jet multiplicity inclusively,
by the particles in hemispheres of 
gluon-gluon ($gg$) systems in an overall color singlet.
Quark jets are defined analogously as hemispheres
of quark-antiquark ($q\overline{q}$) systems.
The hemisphere definition of jets yields results which
are independent of a jet finder.
Therefore, these jets are called ``unbiased.''
Unbiased gluon jet multiplicity has so far been measured 
only in $\Upsilon$~\cite{bib-cleo92,bib-cleo97} and
Z$^0$~\cite{bib-opalgincl96}-\cite{bib-opalgincl98} decays,
corresponding to jet energies of about
5 and 40~GeV, respectively.
In contrast,
unbiased quark jet multiplicity is easy to measure
and has been determined at many scales
(for a recent review, see~\cite{bib-dgphysrep}).

It is of interest to measure
unbiased gluon jet multiplicity at other scales.
Such measurements would allow a test of 
recent predictions~\cite{bib-lupia,bib-capella}
from Quantum Chromodynamics (QCD)
for the scale dependence of the multiplicity 
ratio $r$ between gluon and quark jets and for
the related ratios of slopes $r^{(1)}$ and
of curvatures $r^{(2)}$:
\begin{equation}
  r  \; \equiv  \; \frac{ \ng }{ \nq }  \;\;\;\; ,
  \label{eq-one}
\end{equation}
\begin{equation}
  r^{(1)} \; \equiv \; \frac{ {\mathrm{d}}\ng / 
                        {\mathrm{d}}y }
                      { {\mathrm{d}}\nq / 
                        {\mathrm{d}}y }  \;\;\;\; ,
  \label{eq-two}
\end{equation}
\begin{equation}
  r^{(2)} \; \equiv \; \frac{ {\mathrm{d}}^2\ng / 
                        {\mathrm{d}}^2 y }
                      { {\mathrm{d}}^2\nq / 
                        {\mathrm{d}}^2 y }  \;\;\;\; ,
  \label{eq-three}
\end{equation}
where $\ng$ and $\nq$ are the mean particle multiplicities
of gluon and quark jets,
$y$$\,=\,$$\ln\,(Q/\Lambda)$,
$Q$ is the jet energy and $\Lambda$ is the QCD scale parameter.
Recently,
a method to extract unbiased gluon jet multiplicity
from {\it biased} 
{\epem}$\,\rightarrow\,$$q\overline{q}g$
events was proposed~\cite{bib-eden,bib-edenkhoze},
extending earlier formalism~\cite{bib-dok88}.
This method provides an indirect means to determine
the particle multiplicity of gluon jets,
in a manner which corresponds
to the hemisphere definition of jets.
By combining measurements of $\ng$ found from this method
with the unbiased measurements for $\nq$,
the ratios $r$, $r^{(1)}$ and $r^{(2)}$ can be determined
at a variety of scales and used to test the corresponding
QCD results in a quantitative manner.

The purpose of the present study is to test this 
method~\cite{bib-eden,bib-edenkhoze}
to obtain unbiased measurements of $\ng$
and to apply the results to determine the ratios,
eqs.~(\ref{eq-one})-(\ref{eq-three}).
The data were collected with the OPAL detector operating
at the LEP accelerator at CERN.
We determine the ratios (\ref{eq-one})-(\ref{eq-three})
as a function of scale and compare the results to
recent QCD predictions~\cite{bib-lupia,bib-capella}.
Previously published measurements of the scale dependence 
of $r$~\cite{bib-delphi96,bib-delphi99,bib-opal00}
and $r^{(1)}$~\cite{bib-delphi99,bib-opal00}
were based on biased jet samples,
whereas experimental results for the ratio of 
curvatures~$r^{(2)}$ have not previously been presented.
We also use the results on the scale dependence of
unbiased gluon and quark jet multiplicities to 
derive a measurement of the ratio of QCD color factors
$\caa/\cff$.

\section{Theoretical Framework}
\label{sec-theory}

Analytic expressions for the mean particle multiplicity
of {\epem} three-jet events,
valid to the next-to-leading order of perturbation
theory (NLO, also called MLLA~\cite{bib-mlla}),
were recently presented in~\cite{bib-edenkhoze}:
\begin{eqnarray}
  \nqqg & = & \nqq\,(L,\ktlu) + \frac{1}{2}\, \ngg\,(\ktlu)
 \label{eq-eden14b} \;\;\;\; , \\
  \nqqg & = & \nqq\,(\lqq,\ktlu) + \frac{1}{2}\, \ngg\,(\ktle) 
 \label{eq-eden14a}
   \;\;\;\; .
\end{eqnarray}
These results,
along with the others presented in this section,
are based on massless quarks.
The reason for the two different expressions,
eqs.~(\ref{eq-eden14b}) and~(\ref{eq-eden14a}),
is that there is an ambiguity in the definition of the gluon jet
transverse momentum when the gluon radiation is hard.
$\ktlu$ and $\ktle$ are the transverse momenta of the
gluon with respect to the quark-antiquark system using
the definition of either the Lund ($\ktlu$)~\cite{bib-eden}
or Leningrad ($\ktle$)~\cite{bib-dok88} groups,
see~\cite{bib-edenkhoze} and references therein
for further discussion.
Note that due to QCD coherence,
the multiplicity of a gluon jet produced in {\epem} annihilations 
depends on its transverse momentum and not 
its energy (see for example~\cite{bib-dok88}).
$\nqqg$ is the particle multiplicity of a three-jet event
selected using a jet algorithm (see below).
$\nqq$ and $\ngg$ are the multiplicities of two-jet
$q\overline{q}$ and $gg$ systems,
given by twice $\nq$ and $\ng$, respectively.
The scales $L$, $\lqq$, $\ktlu$ and $\ktle$ 
are defined as follows:
\begin{eqnarray}
   L & = & \ln\left( \frac{ s }{ \Lambda^2 } \right) 
           \label{eq-lll} \;\;\;\; , \\
   \lqq & = & \ln\left( \frac{ \sqq }
      { \Lambda^2 } \right) \label{eq-lqq} \;\;\;\; , \\
   \ktlu & = & 2\,\ln\left( \frac{ {\ptlu} }{ \Lambda } \right)
           \label{eq-ktlu} \;\;\;\; ,  \\
   \ktle & = & 2\,\ln\left( \frac{ {\ptle} }{ \Lambda } \right)
           \label{eq-ktle} \;\;\;\; , \\
%   s & = & \ecm^2 \;\;\;\; , \\
   \sqq & = & p_q \cdot p_{\overline{q}} \label{eq-sqq} \;\;\;\; , \\
   {\ptlu} & = & \sqrt {\frac{ s_{qg}s_{\overline{q}g} }{ s } }
           \label{eq-ptlu} \;\;\;\; , \\
   {\ptle} & = & \sqrt{ \frac{ s_{qg}s_{\overline{q}g} }
                 { \sqq } }
           \label{eq-ptle} \;\;\;\; , 
%   s_{qg} & = & p_q \cdot p_g \;\;\;\; , \\
%   s_{\overline{q}g} & = & p_{\overline{q}} \cdot p_g \;\;\;\; ,
\end{eqnarray}
with $s$$\,=\,$$\ecm^2$, $s_{ig}$$\,=\,$$p_i \cdot p_g$
($i$$\,=\,$$q,\overline{q}$),
and $p_q$, $p_{\overline{q}}$ and $p_g$
the 4-momenta of the $q$, $\overline{q}$ and~$g$.
$L$~specifies the {\epem} c.m. energy ($\ecm$) and
$\lqq$~the energy of the $q\overline{q}$ system
in the $q\overline{q}$ rest frame.

The multiplicity of the $gg$ system in this formalism,
$\ngg$,
depends only on a single scale:
$\ktlu$ in eq.~(\ref{eq-eden14b}) or
$\ktle$ in eq.~(\ref{eq-eden14a}).
This dependence on a single scale is a statement
that $\ngg$ is {\it unbiased},
i.e. $\ngg\,(\kperp)$ in eq.~(\ref{eq-eden14b})
or~(\ref{eq-eden14a}) is equivalent to the inclusive
multiplicity of a $gg$ event from a color singlet source 
produced at the same scale~$\kperp$,
to NLO accuracy.
In contrast,
the multiplicity of the $\qq$ system, $\nqq$,
depends on {\it two} scales:
$L$ and $\ktlu$ in eq.~(\ref{eq-eden14b}) or
$\lqq$ and $\ktlu$ in eq.~(\ref{eq-eden14a}).
This dependence on two scales implies that
the multiplicity of the $\qq$ system is {\it biased},
depending
on a hard scale $L$ or $\lqq$ for the production
of the quark and antiquark jets
and on a cutoff $\kperp$$\,\approx\,$$\ktlu$
below which the gluon jet is not resolved.
This accounting for the bias in quark jet multiplicity
due to the jet finder criteria used to select 
the $q\overline{q}g$ events
is the difference between the theoretical results
used here~\cite{bib-eden}
and those presented previously~\cite{bib-dok88}.

The above expressions are valid for jet algorithms 
which employ a transverse momentum cutoff
$\kperp$ to resolve ``two-jet'' $\qq$ from 
``three-jet'' \mbox{$\qq g$} events.
Examples of such algorithms are the Durham~\cite{bib-durhamjf},
Cambridge~\cite{bib-cambridgejf} and
Luclus~\cite{bib-luclusjf} jet finders.
The formalism is not expected to be valid for
the Jade~\cite{bib-jadejf} or cone jet finders~\cite{bib-conejf},
which use invariant masses or angles rather than transverse momenta
to separate the two- and three-jet event classes.

According to the prescription in~\cite{bib-edenkhoze},
the resolution scale $\kperp$ should be adjusted separately
for each event so exactly three jets are reconstructed.
In contrast,
a fixed value of the resolution scale results in
truncation of higher order radiation,
biasing the gluon jet's properties.
Note that any radiation (``sub-jet'') emitted within
the quark or gluon jet must necessarily have a smaller
transverse momentum than the gluon jet itself:
otherwise the role of the ``gluon jet'' and ``sub-jet''
would be reversed.
Thus the transverse momentum of the gluon jet,
$\ktgluon$,
defines an effective cutoff for sub-jet radiation,
i.e. $\kperp$$\,=\,$$\ktgluon$$\,\approx\,$$\ktlu$.

In the formalism of~\cite{bib-eden},
the biased quark jet multiplicities $\nqq\,(J,\ktlu)$
in eqs.~(\ref{eq-eden14b}) and~(\ref{eq-eden14a})
(with $J$ representing either $L$ or $\lqq$) 
can be obtained from the corresponding {\it unbiased} 
multiplicities $\nqq\,(J^{\prime})$,
which depend only on a single scale~$J^{\prime}$,
through the relation:
\begin{equation}
  \nqq\,(J,\ktlu) \; = \; \nqq\,(J^{\prime})
     + (J - J^{\prime})\, \frac{{\mathrm{d}}\,\nqq\,(J^{\prime}) }
        {{\mathrm{d}} J^{\prime} } \;\;\;\; ,
  \label{eq-qbiased}
\end{equation}
where $J^{\prime}$$\,=\,$$\ktlu+c_q$
with $c_q$$\,=\,$3/2.
The unbiased term $\nqq\,(K)$ is equivalent to the
mean multiplicity of inclusive
{\epem}$\,\rightarrow\,$$hadrons$ events 
as a function of $K$$\,=\,$$\ecm$.

Finally,
a relationship between the scale evolution of unbiased
gluon and quark jet multiplicities is given~\cite{bib-eden}:
\begin{equation}
  \frac{{\mathrm{d}}\,\ngg\,(L^{\prime}) } {{\mathrm{d}} L }
  = \frac{\caa}{\cff} 
      \left(1 - \frac{\alpha_0 c_r}{L} \right)
      \frac{{\mathrm{d}}\,\nqq\,(L) } {{\mathrm{d}} L }
       \;\;\;\; ,
  \label{eq-gluonevolution}    
\end{equation}
where $L^{\prime}$$\,=\,$$L+c_g-c_q$, $c_g$$\,=\,$11/6,
$\alpha_0$$\,=\,$$6\,\caa/(11\,\caa-2\,\nff)$,
$c_r$$\,=\,$$10\,\pi^2/27 - 3/2$,
and $\nff$ equals the number of active quark flavors,
taken to be $\nff$$\,=\,$5.
$\caa$ and $\cff$ are the QCD color factors,
3 and 4/3, respectively.

\section{Detector and data sample}

\label{sec-detector}

The OPAL detector is described in
detail elsewhere~\cite{bib-detector,bib-si}.
The tracking system
consists of a silicon microvertex detector,
an inner vertex chamber,
a large volume jet chamber,
and specialized chambers at the outer radius of the 
jet chamber which improve the measurements in the
$z$-direction.\footnote{Our
coordinate system is defined so that
$z$~is the coordinate parallel to the e$^-$ beam axis,
$r$~is the coordinate normal to the beam axis,
$\phi$~is the azimuthal angle around the beam axis and
$\theta$~is the polar angle \mbox{with respect to~$z$.}}
The tracking system covers the region
$|\cos\theta|$$\,<\,$0.98 and
is enclosed by a solenoidal magnet coil
with an axial field of~0.435~T.
Electromagnetic energy is measured by a
lead-glass calorimeter located outside the magnet coil,
which also covers $|\cos\theta|$$\,<\,$0.98.

The present analysis is based on a sample of
about $2\,283\,000$ hadronic annihilation events,
corresponding to the OPAL sample collected within 0.3~GeV 
of the Z$^0$ peak that includes readout of both the $r$-$\phi$ and $z$ 
coordinates of the silicon strip microvertex detector~\cite{bib-si}.
The procedures for identifying hadronic annihilation
events are described in~\cite{bib-opaltkmh}.

We employ charged tracks measured in the tracking chambers
and clusters measured in the electromagnetic calorimeter.
Charged tracks are required to have at least 20 measured
points (of 159 possible) in the jet chamber,
or at least 50\% of the points expected based on the
track's polar angle,
whichever is smaller.
In addition, the tracks are required 
to have a momentum in the direction perpendicular 
to the beam axis greater than 0.05~GeV/$c$,
to lie in the region $|\cos\theta|$$\,<\,$0.96,
to point to the origin to within 5~cm in the $r$-$\phi$ plane
and 30~cm in the $z$ direction,
and to yield a reasonable $\chi^2$ per
degree-of-freedom (d.o.f.)
for the track fit in the $r$-$\phi$ plane.
Clusters are required to have an energy greater 
than 0.10~GeV if they are in the barrel section 
of the detector ($|\cos\theta|$$\,<\,$0.82)
or 0.25~GeV if they are in the endcap section 
(0.82$\,<\,$$|\cos\theta|$$\,<\,$0.98).
A matching algorithm is employed to
reduce double counting of energy in cases where 
charged tracks point towards electromagnetic clusters.
Specifically,
the expected calorimeter energy of the associated
tracks is subtracted from the cluster energy.
If the energy of a cluster is smaller than 
that expected for the associated tracks,
the cluster is not used.
Each accepted track and cluster
is considered to be a particle.
Tracks are assigned the pion mass.
Clusters are assigned zero mass since they originate
mostly from photons.

To eliminate residual background and events
in which a significant number of particles is lost
near the beam direction,
the number of accepted charged tracks in an event
is required to be at least five and the 
thrust axis~\cite{bib-thrust} of the event,
calculated using the particles,
is required to satisfy
$|\cos (\theta_{\mathrm{thrust}})|$$\,<\,$0.90,
where $\theta_{\mathrm{thrust}}$ is the
angle between the thrust and beam axes.
The residual background to the sample of hadronic events
from all sources is estimated to be less 
than~1\%~\cite{bib-opaltkmh}.

\section{Selection of uds events}
\label{sec-uds}

The theoretical formalism we employ is based on massless quarks,
as stated in Sect.~\ref{sec-theory}.
Therefore,
beginning with the data sample described in Sect.~\ref{sec-detector},
we select light quark (u, d and~s) events for our study.

The uds event tagging is based on 
the signed impact parameter values of charged tracks
with respect to the primary event vertex, $\dsign$.
The magnitude of $\dsign$ is given by the distance of closest
approach of the track to the event vertex
in the $r$-$\phi$ plane.
The sign of $\dsign$ is determined as follows.
Jets are reconstructed in each event using the
cone jet finder~\cite{bib-conejf}.\footnote{The 
jet resolution parameters used are:
a cone half angle of 0.55~radians and 
a minimum jet energy of 5~GeV.}
%Jade jet finder~\cite{bib-jadejf}.
%Following a procedure developed by OPAL for B hadron physics,
%$y_{cut}$ is allowed to vary from event-to-event as
%$y_{cut}=(49\;{\mathrm GeV}/\evis)^2$,
%where $\evis$ is the sum of particle energy in the
%event and the value of 49~GeV was chosen to optimize 
%momentum resolution in B hadron decays.
For each track assigned to a jet by the jet finder,
hemispheres are defined by the plane 
perpendicular to the corresponding jet axis.
The sign of $\dsign$ is determined using the crossing point
in the $r$-$\phi$ plane between the track and 
jet axis to which it is assigned.
$\dsign$ is positive if the crossing point is 
in the forward hemisphere and negative otherwise.
The forward hemisphere is the hemisphere 
which contains the jet.
Because of the relatively long lifetime of 
hadrons containing c and b quarks,
the distribution of $\dsign$
is strongly skewed toward positive values 
for c and b events but not for uds events.

Note that the jet finding procedure described in the
previous paragraph is employed solely for the
purpose of identifying uds events and not for
the analysis of gluon jet multiplicity presented below in
Sects.~\ref{sec-procedure}-\ref{sec-results}.

Charged tracks are selected for the uds tagging procedure
if they have $r$-$\phi$ coordinate information 
from at least one of the two silicon detector layers,
a momentum of 1~GeV/$c$ or larger,
and a maximum distance of closest approach
to the primary event vertex in the $r$-$\phi$ plane of 0.3~cm
with a maximum uncertainty on this quantity of~0.1~cm.
In addition they must be assigned to a jet using the
cone jet finding procedure mentioned above.
If no track in an event satisfies these requirements
(1.5\% of the events), the event is eliminated.
The number of tracks which meet these requirements and 
have $\dsign/\sigma_{\dsign}$$\,>\,$3.0 is determined,
where $\sigma_{\dsign}$ is the uncertainty associated with~$\dsign$.
An event is tagged as a uds event if this number is zero.
In total $1\,109\,017$ events are tagged.
The number of events in our final event sample
and the corresponding uds purity
are presented in Sect.~\ref{sec-procedure}.

\section{Analysis method}
\label{sec-procedure}

Three-jet events are defined using a jet algorithm.
For our standard analysis
we employ the Durham jet finder~\cite{bib-durhamjf}.
As a systematic check,
we use the Cambridge~\cite{bib-cambridgejf}
and Luclus~\cite{bib-luclusjf} jet finders
(see Sect.~\ref{sec-systematic}).
We choose these algorithms because they employ a cutoff 
in transverse momentum to separate two- and three-jet events,
in correspondence to the theoretical formalism 
of~\cite{bib-eden,bib-edenkhoze}.
In particular,
this means that eqs.~(\ref{eq-eden14b}), (\ref{eq-eden14a}),
(\ref{eq-qbiased}) and~(\ref{eq-gluonevolution})
are expected to be applicable for these jet algorithms.

\begin{figure}[tp]
\begin{center}
  \epsfxsize=8cm
  \epsffile{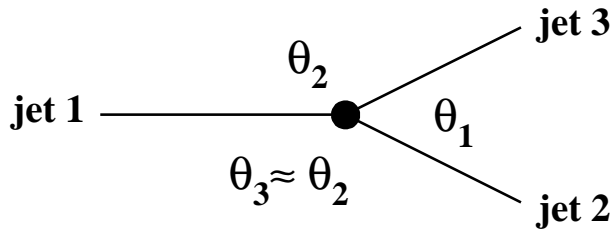}
\end{center}
\caption{
Schematic representation of a three-jet 
quark-antiquark-gluon $q\overline{q}g$ event
with a Y event topology~\cite{bib-opalse91,bib-opalqg91},
in which the angle between the highest energy jet
and each of the two lower energy jets is about the same.
The jets are ordered in energy such that jet~1 has
the highest energy.
The angle $\theta_1$ opposite the highest energy jet
is used to specify the event topology.
In our analysis,
the gluon jet is assumed to be jet~3.
}
\label{fig-yevent}
\end{figure}

The resolution scale of the Durham jet finder, $\ycut$,
is adjusted separately for each tagged uds event so
exactly three jets are reconstructed.
For the Cambridge jet finder,
the resolution scale is again~$\ycut$.
For Luclus, the corresponding parameter is~$d_{join}$.
We apply this procedure rather than use
a fixed resolution scale to avoid introducing
a bias in the gluon jet properties as discussed in
Sect.~\ref{sec-theory}.
The jets are assigned energies based on the angles
between them,
assuming massless kinematics
(see for example~\cite{bib-opalse91}).
The jets are ordered from 1 to 3
such that jet~1 has the highest energy.
Events are retained if the angles between the highest
energy jet and the other two are the same to within~3$^\circ$,
the so-called ``Y events.''\footnote{Y events were first
studied in~\cite{bib-opalse91,bib-opalqg91};
the overall charged multiplicity of Y events was first
presented in~\cite{bib-delphi99}.}
A schematic diagram of a Y event is shown in Fig.~\ref{fig-yevent}.
For Y events,
the scales $\lqq$, $\ktlu$ and $\ktle$
(eqs.~(\ref{eq-lqq})-(\ref{eq-ktle}))
depend only on $\ecm$ and one inter-jet angle,
conveniently chosen to be~$\theta_1$
(see Fig.~\ref{fig-yevent}).
For $35^{\circ}$$\,\leq\,$$\theta_1$$\,\leq\,$$120^{\circ}$,
the range of $\theta_1$ we employ for our 
gluon jet analysis
(Sect.~\ref{sec-gluonmult}),
$22\,365$ events are selected:
this is our final event sample.
We require $\theta_1$$\,\geq\,$$35^{\circ}$ in order that 
the two lower energy jets be clearly separated from
each other.
Note that since $\theta_2$ and $\theta_3$ (see Fig.~\ref{fig-yevent})
are not exactly equal,
the energies of jets~2 and~3 are not exactly the same.

The uds event purity of this sample is estimated 
using the Jetset Monte Carlo 
multihadronic event generator~\cite{bib-jetset}
including initial state photon radiation,
detector simulation~\cite{bib-gopal},
and the same analysis
procedures as are applied to the data.
We use a sample of about $6\,000\,000$ Monte Carlo events
generated with version~7.4 of the program
and the parameters given in~\cite{bib-qg95b}.
The estimated uds purity
is found to be~$(78.5\pm 0.2\,\mathrm{(stat.)})$\%.
The Monte Carlo predicts that 70\% of the background
events are c events and 30\% are b events.
The ratio of the number of events 
in the final sample to the number in the initial
multihadronic Z$^0$ decay sample is
$(0.980\pm 0.007\,\mathrm{(stat.)})\times 10^{-2}$
for the data and 
$(1.036\pm 0.004\,\mathrm{(stat.)})\times 10^{-2}$
for the Monte Carlo.

For simplicity,
we identify the gluon jet by assuming it is the 
lowest energy jet in an event, i.e. jet~3.
For the majority of Y events,
$\theta_1$$\,<<\,$$\theta_2, \theta_3$.
For these events,
jet~1 has an overwhelming probability (typically about 97\%)
to be a quark (or antiquark) jet,
while jets~2 and~3 have about an equal probability
to arise from a quark or gluon.
Since $\theta_2$$\,\approx\,$$\theta_3$
(see Fig.~\ref{fig-yevent}),
the scales $\lqq$, $\ktlu$ and $\ktle$ are
almost unchanged if the gluon is actually jet~2
rather than jet~3.
%not very sensitive to whether the gluon is jet~2 or~3
Similarly,
as the energy of jet~3 increases so that
$\theta_1$$\,\approx\,$$\theta_2$$\,\approx\,$$\theta_3$,
all three jets have about the same probability to
be the gluon jet and the determination of scales
is again insensitive to whether the 
gluon jet assignment is correct or not.
This relative insensitivity to gluon jet misidentification
is our motivation for employing Y events.

The scales $\lqq$, $\ktlu$ and $\ktle$ and the 
three-jet event charged particle multiplicity $\nqqgch$
are measured as a function of~$\theta_1$.
$\lqq$, $\ktlu$ and $\ktle$
are determined using all accepted charged 
and neutral particles.
For the QCD scale parameter $\Lambda$,
used in the definition of these last three variables
(also for the definition of $L$, see eq.~(\ref{eq-lll})),
we choose a value of 0.20~GeV.
We verified our results are not sensitive to this choice.

The measured distributions of
$\nqqgch$, $\lqq$, $\ktlu$ and $\ktle$
are corrected for experimental acceptance and resolution,
the effects of initial state photon radiation,
and misidentification of uds events,
using bin-by-bin 
multiplicative factors derived from Jetset.
The corrections for $\lqq$, $\ktlu$ and $\ktle$
also account for gluon jet misidentification,
i.e. for the contributions of
events in which the gluon jet is not jet~3.
The data are corrected to the hadron level.
The hadron level does not include detector response
or initial state radiation and treats
all charged and neutral particles with lifetimes greater
than \mbox{$3\times 10^{-10}$~s} as stable:
hence charged particles from the decays of K$_{\mathrm S}^0$
and weakly decaying hyperons are included in the
corrected distributions.
The correction factors are
determined as described in~\cite{bib-opales90}.
The corrections are typically 1.08 for $\nqqgch$,
0.98 for~$\lqq$, 1.03 for $\ktlu$, and 1.05 for~$\ktle$,
and do not vary much with~$\theta_1$.
There is good agreement between the distributions of the
uncorrected data and the Monte Carlo sample when the latter 
includes initial-state radiation and detector simulation,
with the same analysis procedures
applied to the simulated events as to the data.

\begin{figure}[tp]
\begin{center}
  \epsfxsize=15cm
  \epsffile{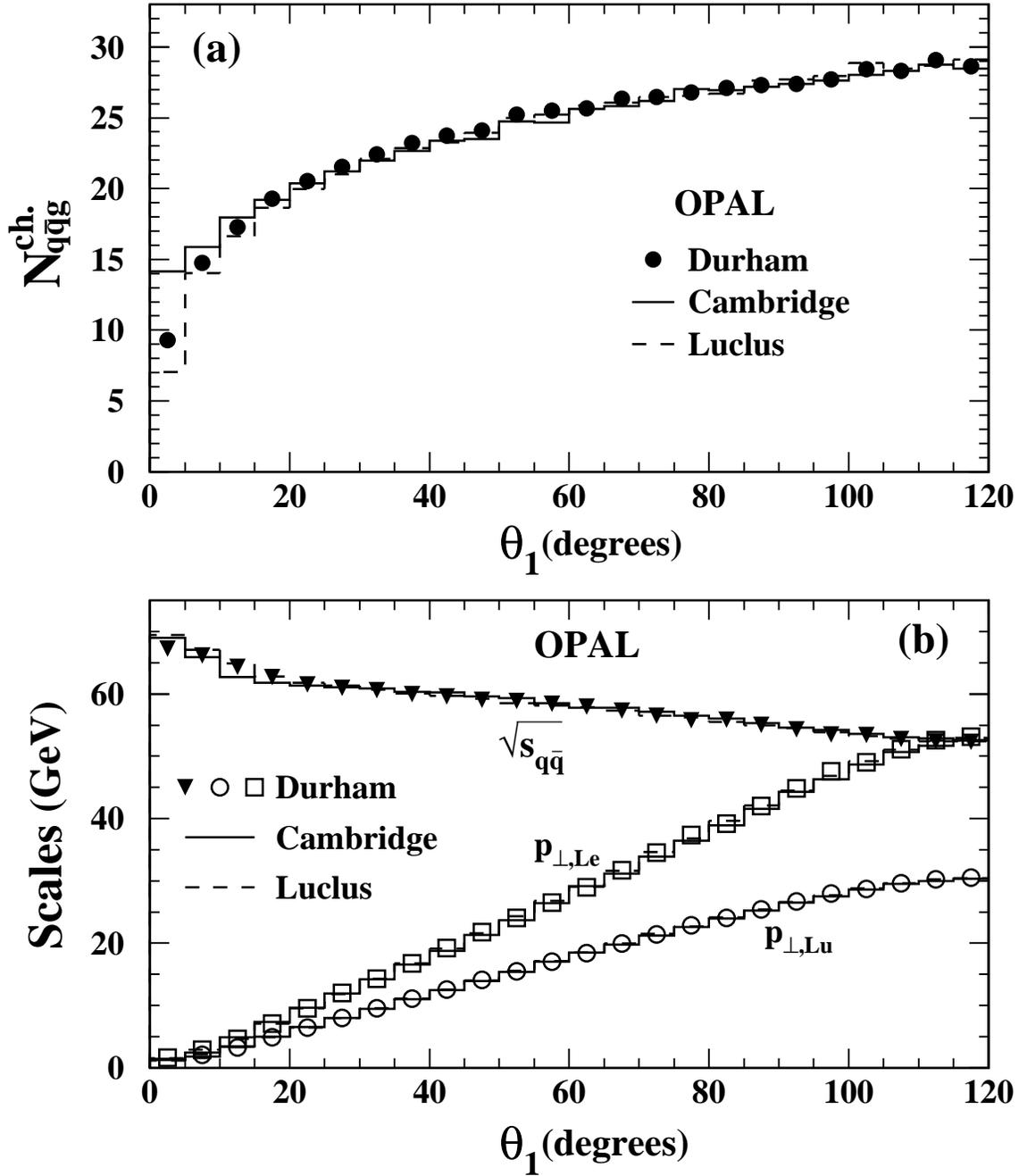}
\end{center}
\caption{
(a)~The mean charged particle multiplicity of
three-jet uds flavor Y events from Z$^0$ decays,
selected using the Durham~\cite{bib-durhamjf},
Cambridge~\cite{bib-cambridgejf}
and Luclus~\cite{bib-luclusjf} jet finders,
as a function of the opening angle~$\theta_1$;
(b)~the corresponding scales $\sqrt{\sqq}$, $\ptlu$ and $\ptle$,
see eqs.~(\ref{eq-sqq})-(\ref{eq-ptle}).
The data have been corrected for initial state photon radiation,
detector response,
misidentification of uds events,
and (for part~(b)) misidentification of gluon jets.
The uncertainties (too small to be visible) 
are statistical only.
}
\label{fig-scales}
\end{figure}

\begin{table}[t]
\centering
\begin{tabular}{|c|cccc|}
 \hline
 & & & & \\[-.2cm]
 $\theta_{1}$ (degrees) & $\nqqgch$ & $\sqrt{s_{q\bar{q}}}$ (GeV)
     & $p_{\perp,Lu}$ (GeV) &$p_{\perp,Le}$ (GeV)  \\[.3cm]
\hline
 & & & & \\[-.2cm]
0-5     &$9.31\pm0.13$    &$67.27\pm0.18$   &$1.227\pm0.007$  &$1.595\pm0.016$\\
5-10    &$14.739\pm0.026$ &$66.182\pm0.023$ &$2.104\pm0.003$  &$2.897\pm0.005$ \\
10-15   &$17.279\pm0.024$ &$64.385\pm0.024$ &$3.291\pm0.003$  &$4.658\pm0.005$\\
15-20   &$19.282\pm0.038$ &$62.679\pm0.048$ &$4.885\pm0.005$  &$7.102\pm0.012$ \\
20-25   &$20.540\pm0.056$ &$61.530\pm0.070$ &$6.469\pm0.009$  &$9.581\pm0.023$  \\
25-30   &$21.533\pm0.073$ &$60.945\pm0.086$ &$8.008\pm0.013$  &$11.975\pm0.035$ \\
30-35   &$22.417\pm0.092$ &$60.597\pm0.097$ &$9.518\pm0.017$  &$14.315\pm0.048$ \\
35-40   &$23.23\pm0.11$   &$60.02\pm0.11$   &$11.057\pm0.022$ &$16.790\pm0.064$  \\
40-45   &$23.75\pm0.13$   &$59.61\pm0.12$   &$12.553\pm0.027$ &$19.194\pm0.079$ \\
45-50   &$24.12\pm0.15$   &$59.04\pm0.13$   &$14.090\pm0.033$ &$21.751\pm0.098$  \\
50-55   &$25.24\pm0.17$   &$58.88\pm0.14$   &$15.514\pm0.038$ &$24.01\pm0.11$  \\
55-60   &$25.49\pm0.19$   &$58.45\pm0.14$   &$16.991\pm0.043$ &$26.49\pm0.13$  \\
60-65   &$25.66\pm0.21$   &$57.94\pm0.15$   &$18.423\pm0.049$ &$28.98\pm0.15$  \\
65-70   &$26.34\pm0.23$   &$57.32\pm0.15$   &$19.922\pm0.055$ &$31.68\pm0.17$  \\
70-75   &$26.47\pm0.25$   &$56.55\pm0.16$   &$21.443\pm0.061$ &$34.56\pm0.19$  \\
75-80   &$26.79\pm0.26$   &$55.81\pm0.16$   &$22.879\pm0.065$ &$37.36\pm0.21$  \\
80-85   &$27.10\pm0.28$   &$55.89\pm0.14$   &$24.018\pm0.064$ &$39.21\pm0.20$  \\
85-90   &$27.29\pm0.31$   &$55.04\pm0.14$   &$25.388\pm0.065$ &$42.04\pm0.21$  \\
90-95   &$27.37\pm0.31$   &$54.29\pm0.13$   &$26.687\pm0.066$ &$44.80\pm0.22$  \\
95-100  &$27.72\pm0.33$   &$53.48\pm0.12$   &$27.927\pm0.065$ &$47.59\pm0.22$  \\
100-105 &$28.42\pm0.36$   &$53.384\pm0.097$ &$28.703\pm0.054$ &$49.00\pm0.18$  \\
105-110 &$28.31\pm0.35$   &$52.779\pm0.077$ &$29.620\pm0.046$ &$51.15\pm0.15$  \\
110-115 &$29.08\pm0.36$   &$52.416\pm0.053$ &$30.240\pm0.032$ &$52.58\pm0.11$  \\
115-120 &$28.62\pm0.37$   &$52.328\pm0.032$ &$30.497\pm0.020$ &$53.115\pm0.066$ 
    \\[.2cm]
\hline
\end{tabular}
\caption{Measurements of the mean charged particle multiplicity $\nqqgch$
of three-jet uds flavor Y events from Z$^0$ decays,
as a function of the
angle $\theta_1$ between the two lowest energy jets.
The jets are defined using the Durham jet-finder.
The results for the quark jet scale $\sqrt{\sqq}$
and the gluon jet scales $\ptlu$ and $\ptle$ are also given.
The uncertainties are statistical only.
These data are shown in Fig.~\ref{fig-scales}.
}
 \label{tab-scales}
\end{table} 

Corrected distributions of $\nqqgch$, $\sqrt{\sqq}$,
$\ptlu$ and $\ptle$ are shown in Fig.~\ref{fig-scales}.
The last three variables are obtained from the corrected
distributions of $\lqq$, $\ktlu$ and $\ktle$
using eqs.~(\ref{eq-lqq})-(\ref{eq-ktle}).
We prefer to present $\sqrt{\sqq}$,
$\ptlu$ and $\ptle$ rather than $\lqq$, $\ktlu$ and $\ktle$
because they provide a clearer illustration of
the physical scales in our analysis.
These data are listed in Table~\ref{tab-scales}.
Besides the results from our standard analysis,
the results found using the Cambridge and Luclus jet
finders are shown in Fig.~\ref{fig-scales}.
For about 2\% of the events,
the resolution parameter $y_{cut}$ of the
Cambridge algorithm could not be adjusted to
yield three reconstructed jets,
see~\cite{bib-stan} for a discussion of this
feature of the algorithm.
These events were rejected for the Cambridge jet
finder based study.

We invert eqs.~(\ref{eq-eden14b}) and~(\ref{eq-eden14a})
to obtain expressions for the unbiased gluon jet 
charged particle multiplicities $\nggch\,(\kperp)$:
\begin{eqnarray}
  \nggch\,\left( \ktlu\,(\theta_1) \right) & = &
    2\,\left[ \nqqgch\,(L,\theta_1) - 
       \nqqch \left( L,\ktlu\,(\theta_1) \right) \right]
   \label{eq-nggtwo} \;\;\;\; , \\
  \nggch\,\left( \ktle\,(\theta_1) \right) & = &
    2\,\left[ \nqqgch\,(L,\theta_1) - 
       \nqqch \left( \lqq\,(\theta_1),\ktlu\,(\theta_1)
           \right) \right]
       \;\;\;\; ,   \label{eq-nggone}
\end{eqnarray}
where $\nqqgch$, $\lqq$, $\ktlu$ and $\ktle$ are determined
from the Y events as described above.

To find the $\nqqch\,(L,\ktlu)$ terms in eq.~(\ref{eq-nggtwo}),
we employ two methods.
First, 
for the standard analysis,
we perform a direct measurement.
Specifically we determine the particle multiplicity of two-jet 
uds events from Z$^0$ decays 
(i.e. at a fixed value of~$L$ corresponding to $\ecm$$\,=\,$91.2~GeV)
as a function of the jet resolution scale~$\ktlu$.
This result is presented in Sect.~\ref{sec-nqqchbiased}.
Second, 
as a systematic check (see Sect.~\ref{sec-systematic}),
we evaluate the analytic expression,
eq.~(\ref{eq-qbiased}),
as explained in the following paragraph.
To find the $\nqqch\,(\lqq,\ktlu)$ terms
in eq.~(\ref{eq-nggone}),
we utilize only the second of these methods,
i.e. the one based on eq.~(\ref{eq-qbiased}).
A direct measurement is not straightforward in this case 
because $\lqq$, unlike $L$, is a variable quantity.
A direct measurement of $\nqqch\,(\lqq,\ktlu)$ at the scales
relevant for our analysis would in principle require a 
determination of the particle multiplicity in two-jet events 
as a function of $\ptlu$
for c.m. energies below the Z$^0$.
So far, no such measurements are available.

\begin{table}[t]
\centering
\begin{tabular}{|c|ccc|}
\hline
      & & & \\[-.2cm]
      & $\Lambda_{eff.}$ (GeV) & Normalization $K$ & $\chi^2$ (d.o.f.) \\[.2cm]
\hline
      & & & \\[-.2cm]
  $\nqqch\,(L)$, uds quark jets & $0.190\pm0.032$  & $0.136\pm0.009$ 
      & 6.1 (16) \\[.2cm]
  $\nggch\,(L)$, gluon jets     & $0.600\pm0.059$  & $0.201\pm0.011$ 
      & 28.1 (17) \\[.2cm]
\hline
\end{tabular}
\caption{
Results of fits of the 3NLO expressions for quark and
gluon jet multiplicities to measurements of
charged particle multiplicity in unbiased $q\overline{q}$ 
and $gg$ events,
see the solid curves in Figs.~\ref{fig-nchepem}
and~\ref{fig-ggmult}a.
The fits are performed assuming five active quark flavors.
The uncertainties are statistical only.
The $\chi^2$ per degree-of-freedom (d.o.f.) of the fits are
also given.
The $\chi^2$ values are determined using the statistical
uncertainties of the data and not the systematic terms.
}
\label{tab-fitpar}
\end{table}

\begin{figure}[t]
\begin{center}
  \epsfxsize=14cm
  \epsffile{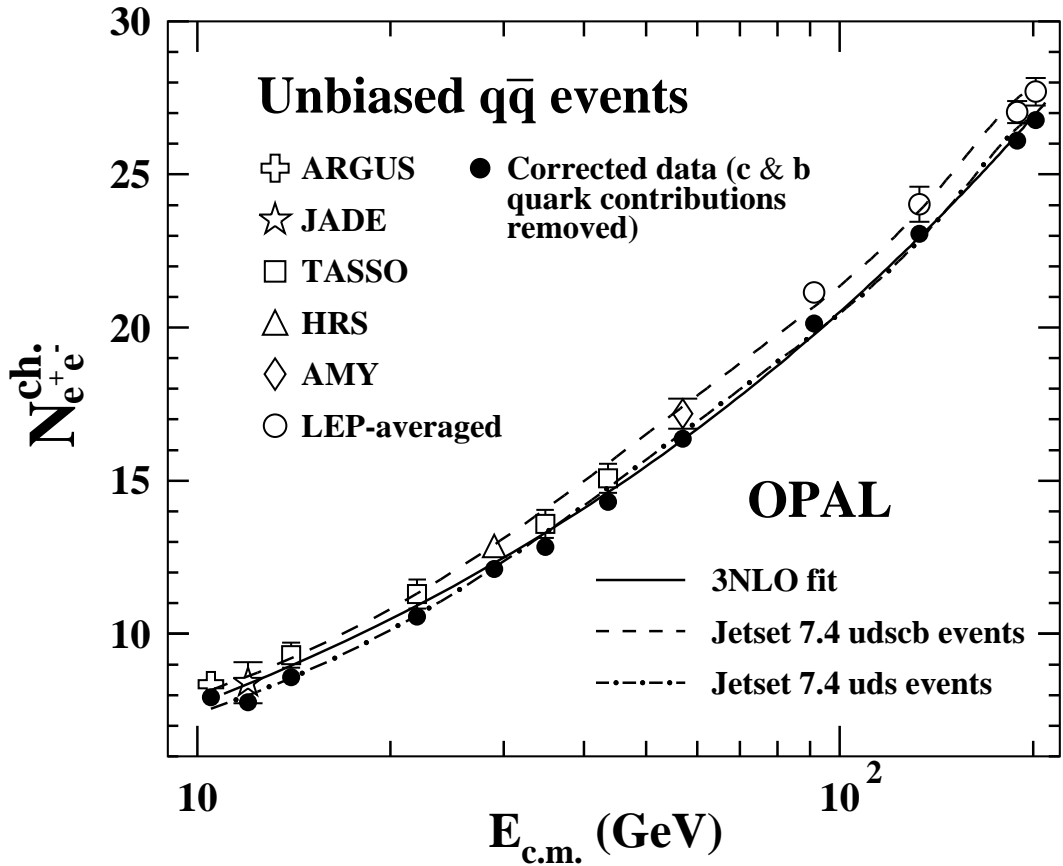}
\end{center}
\caption{
Measurements \protect{(see~\cite{bib-dgphysrep} for
the original references)}
of the inclusive charged particle multiplicity
in {\epem} annihilations (open symbols) as a 
function of~$\ecm$.
The LEP results are averages over the
LEP experiments (see~\cite{bib-dgphysrep}).
The solid points show the corresponding results after
corrections to remove the contributions of c and b events.
For simplicity the uncertainties of the corrected data are
not shown:
these are essentially the same as those of the corresponding
uncorrected data points.
The solid curve is a fit of the next-to-next-to-next-to-leading 
order (3NLO) expression for quark jet 
multiplicity~\cite{bib-capella,bib-dgaryplb}
to the corrected data.
The predictions of the Jetset event generator for udscb and uds
flavor events are also shown.
}
\label{fig-nchepem}
\end{figure}

Our procedure to determine $\nqqch\,(J,\ktlu)$ in 
eqs.~(\ref{eq-nggtwo}) and~(\ref{eq-nggone})
using the analytic expression,
eq.~(\ref{eq-qbiased}),
is as follows.
We begin with the compilation of the mean charged particle
multiplicity of {\epem} hadronic annihilations,
$\nchee$ versus $\ecm$,
given in~\cite{bib-dgphysrep}.
These data are shown as a function of $\ecm$
by the open symbols in Fig.~\ref{fig-nchepem}.
(For simplicity, not all the LEP-averaged
results compiled in~\cite{bib-dgphysrep}
are displayed in Fig.~\ref{fig-nchepem};
however, these data are included in the procedure
described below.)
We correct these results for the contributions of c and b events
so they correspond to our treatment of $\nqqgch$
as explained in Sect.~\ref{sec-uds},
i.e.~to uds events only.
This correction is a bin-by-bin multiplicative
factor determined from Jetset,
given by the ratio of $\nchee$ between uds 
and udscb events as a function of~$\ecm$.
Jetset has been found to provide an accurate description of
the multiplicity of both the uds flavor
and flavor inclusive samples in {\epem} hadronic annihilations 
(see for example~\cite{bib-opalgincl96,bib-opalgincl97,bib-lep2nch}),
justifying its use for this correction.
The predictions of Jetset for $\nchee$ in udscb and uds flavor
events are shown by the dashed and dash-dotted curves
in Fig.~\ref{fig-nchepem}.
The correction factors have values between 0.92 and 0.96.
The corrected results for $\nchee$,
corresponding to the terms
$\nqq\,(J^{\prime})$ in eq.~(\ref{eq-qbiased}) and which we
denote $\nqqch(L)$ in the following,
are shown by the solid points in Fig.~\ref{fig-nchepem}.
We fit these corrected data using the 
next-to-next-to-next-to-leading order
(3NLO) analytic expression for the scale dependence of quark jet
multiplicity $\nq$~\cite{bib-capella,bib-dgaryplb}
assuming $\nff$$\,=\,$5.
Two parameters are fitted:
an effective QCD scale parameter, $\Lambda_{eff.}$,
and an overall normalization constant,~$K$.
The fit results are given in the first
row of Table~\ref{tab-fitpar}.
The result of the fit is shown by the solid curve
in Fig.~\ref{fig-nchepem}.
Using this fitted curve and its derivative,
we numerically evaluate eq.~(\ref{eq-qbiased}) as a function
of $\theta_1$ to determine the biased quark
terms $\nqqch\,(J,\ktlu)$ 
in eqs.~(\ref{eq-nggtwo}) and~(\ref{eq-nggone}).
The derivative is evaluated by differentiating the
analytic expression for $\nq$ with respect to~$y$$\,=\,$$L/2$
and using the fitted parameter values for~$\nqqch(L)$
from Table~\ref{tab-fitpar}.

\section{Results}
\label{sec-results}

\subsection{ The biased quark jet multiplicities
{\boldmath $\nqqch\,(J,\ktlu)$} }
\label{sec-nqqchbiased}

In this section
we present our direct measurement of the biased quark
jet terms $\nqqch\,(L,\ktlu)$ in eq.~(\ref{eq-nggtwo}).
We then use these data to test the analytic expression, 
eq.~(\ref{eq-qbiased}).

\begin{figure}[t]
\begin{center}
  \epsfxsize=14cm
  \epsffile{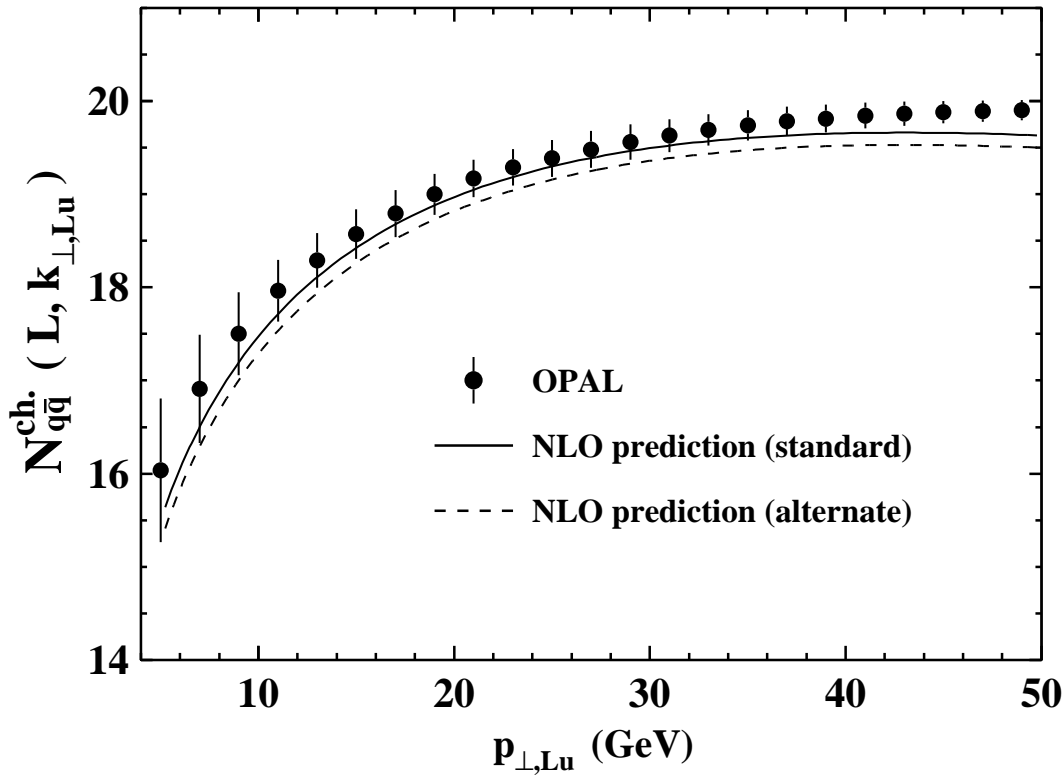}
\end{center}
\caption{
The mean charged particle multiplicity of biased two-jet 
uds flavor events from Z$^0$ decays selected
using the Durham jet finder.
The results are presented as a function of $\ptlu$,
related to the resolution parameter $y_{cut}$
of the jet finder
by $\ptlu$$\,=\,$$E_{vis.}\,\sqrt{y_{cut}}$
with $E_{vis.}$ the event energy.
The data have been corrected for initial state photon radiation,
detector response,
and misidentification of uds events.
The uncertainties are the statistical 
and systematic terms added in quadrature.
The statistical uncertainties are smaller than
the symbol sizes.
The solid curve shows the prediction~\cite{bib-eden} 
of the next-to-leading order (NLO) expression,
eq.~(\ref{eq-qbiased}).
The prediction is seem to agree with the data up to scales
of about 40~GeV.
The dashed curve is the NLO result found
by altering the choice of data points used in the fit
of $\nchee$ versus scale.
}
\label{fig-nqqbiased}
\end{figure}

\begin{table}[tbhp]
\centering
\begin{tabular}{|c|c||c|c|}
\hline
  & & & \\[-.2cm]
 $p_{\perp,Lu}$ (GeV) &  $N^{ch.}_{q\bar{q}}(L,k_{\perp,Lu})$ &
 $p_{\perp,Lu}$ (GeV) &  $N^{ch.}_{q\bar{q}}(L,k_{\perp,Lu})$ \\[.2cm]
\hline
  & & & \\[-.2cm]
 5.0  & $16.04\pm0.04\pm0.77$ &  29.0 & $19.56\pm0.04\pm0.19$ \\
 7.0  & $16.91\pm0.04\pm0.58$ &  31.0 & $19.63\pm0.04\pm0.17$ \\
 9.0  & $17.50\pm0.04\pm0.44$ &  33.0 & $19.69\pm0.04\pm0.16$ \\
 11.0 & $17.96\pm0.04\pm0.33$ &  35.0 & $19.74\pm0.04\pm0.16$ \\
 13.0 & $18.29\pm0.04\pm0.29$ &  37.0 & $19.78\pm0.04\pm0.15$ \\
 15.0 & $18.57\pm0.04\pm0.26$ &  39.0 & $19.81\pm0.04\pm0.14$ \\
 17.0 & $18.79\pm0.04\pm0.25$ &  41.0 & $19.84\pm0.04\pm0.13$ \\
 19.0 & $19.00\pm0.04\pm0.21$ &  43.0 & $19.86\pm0.04\pm0.13$ \\
 21.0 & $19.17\pm0.04\pm0.20$ &  45.0 & $19.88\pm0.04\pm0.12$ \\
 23.0 & $19.29\pm0.04\pm0.19$ &  47.0 & $19.89\pm0.04\pm0.11$ \\
 25.0 & $19.38\pm0.04\pm0.19$ &  49.0 & $19.90\pm0.04\pm0.10$ \\
 27.0 & $19.48\pm0.04\pm0.19$ & & \\[.2cm]
\hline
\end{tabular}
\caption{
Measurements of the mean charged particle multiplicity 
of biased two-jet uds flavor events from Z$^0$ decays
as a function of the transverse momentum cutoff $p_{\perp,Lu}$
used to separate two- and three-jet events.
The jets are defined using the Durham jet-finder.
The first uncertainty is statistical 
and the second systematic.
These data are shown in Fig.~\ref{fig-nqqbiased}.
}
\label{tab-twojet}
\end{table} 

We apply the Durham jet finder to the sample of uds flavor
events described in Sect.~\ref{sec-uds} using a fixed
value of the resolution parameter~$\ycut$ (see below).
Events are retained if exactly two jets are reconstructed.
The solid points in Fig.~\ref{fig-nqqbiased} show 
the charged particle multiplicity of the selected two-jet events
for different choices of~$y_{cut}$.
We relate $y_{cut}$ to the transverse momentum $\ptlu$ through 
$\ptlu$$\,=\,$$E_{vis.}\,\sqrt{y_{cut}}$~\cite{bib-edenkhoze},
with $\evis$ the sum of the particle energy in the event.
The data are corrected for initial state radiation,
the effects of the detector,
and uds event misidentification,
and are plotted as a function of~$\ptlu$.
The solid points in Fig.~\ref{fig-nqqbiased}
therefore represent the direct measurements 
of the biased quark jet terms $\nqqch\,(L,\ktlu)$
mentioned in Sect.~\ref{sec-procedure}.
The vertical lines show the total uncertainties,
with statistical and systematic terms added in quadrature.
The evaluation of systematic uncertainties is
discussed in Sect.~\ref{sec-systematic}.
The statistical uncertainties are negligible compared
to the systematic terms.
These data are listed in Table~\ref{tab-twojet}.
As part of our evaluation of systematic uncertainties,
results analogous to those shown in Fig.~\ref{fig-nqqbiased} 
are determined using the Cambridge jet finder.
For the Cambridge algorithm,
the resolution parameter $y_{cut}$ is
related to $\ptlu$ by the expression given above.
We do not determine analogous results using the Luclus jet finder
because it is not obvious how to relate the resolution
parameter $d_{join}$ to ~$\ptlu$ in this case.

The solid curve in Fig.~\ref{fig-nqqbiased} shows the prediction 
for $\nqqch\,(L,\ktlu)$ found using eq.~(\ref{eq-qbiased}),
evaluated as explained in Sect.~\ref{sec-procedure}.
The curve is seen to describe the direct measurements
fairly well.
In particular,
the shape of the analytic curve is generally similar
to the data,
at least for values of scale below about 40~GeV.
This suggests that the NLO result,
eq.~(\ref{eq-qbiased}),
is adequate for the purposes of our study.
We return to this question in Sect~\ref{sec-gluonmult}.

The difference between the results found using the
Durham and Cambridge jet finders is the dominant
source of systematic uncertainty for
the data in Fig.~\ref{fig-nqqbiased}
(Sect.~\ref{sec-systematic}).
The results from the Durham algorithm 
(standard analysis) lie closer to the NLO curve
than those from the Cambridge algorithm,
especially for small values of~$\theta_1$.

As a systematic check of the NLO prediction for $\nqqch\,(L,\ktlu)$,
we re-evaluated the prediction
after excluding the ARGUS and 91~GeV LEP measurements 
from the fit of $\nchee$ versus~$\ecm$
(see Fig.~\ref{fig-nchepem}).
More details are given in Sect.~\ref{sec-systematic}.
This alternate result is shown by the dashed curve
in Fig.~\ref{fig-nqqbiased}.
The difference between the solid and dashed curves is
seen to be similar to the difference between the
solid curve and symbols.
This suggests that the agreement of the analytic expression 
eq.~(\ref{eq-qbiased}) with the direct measurements
is quite reasonable,
once sources of systematic
uncertainty are considered.

\subsection{ Unbiased gluon jet multiplicity versus scale }
\label{sec-gluonmult}

\begin{figure}[tp]
\begin{center}
  \epsfxsize=15cm
  \epsffile{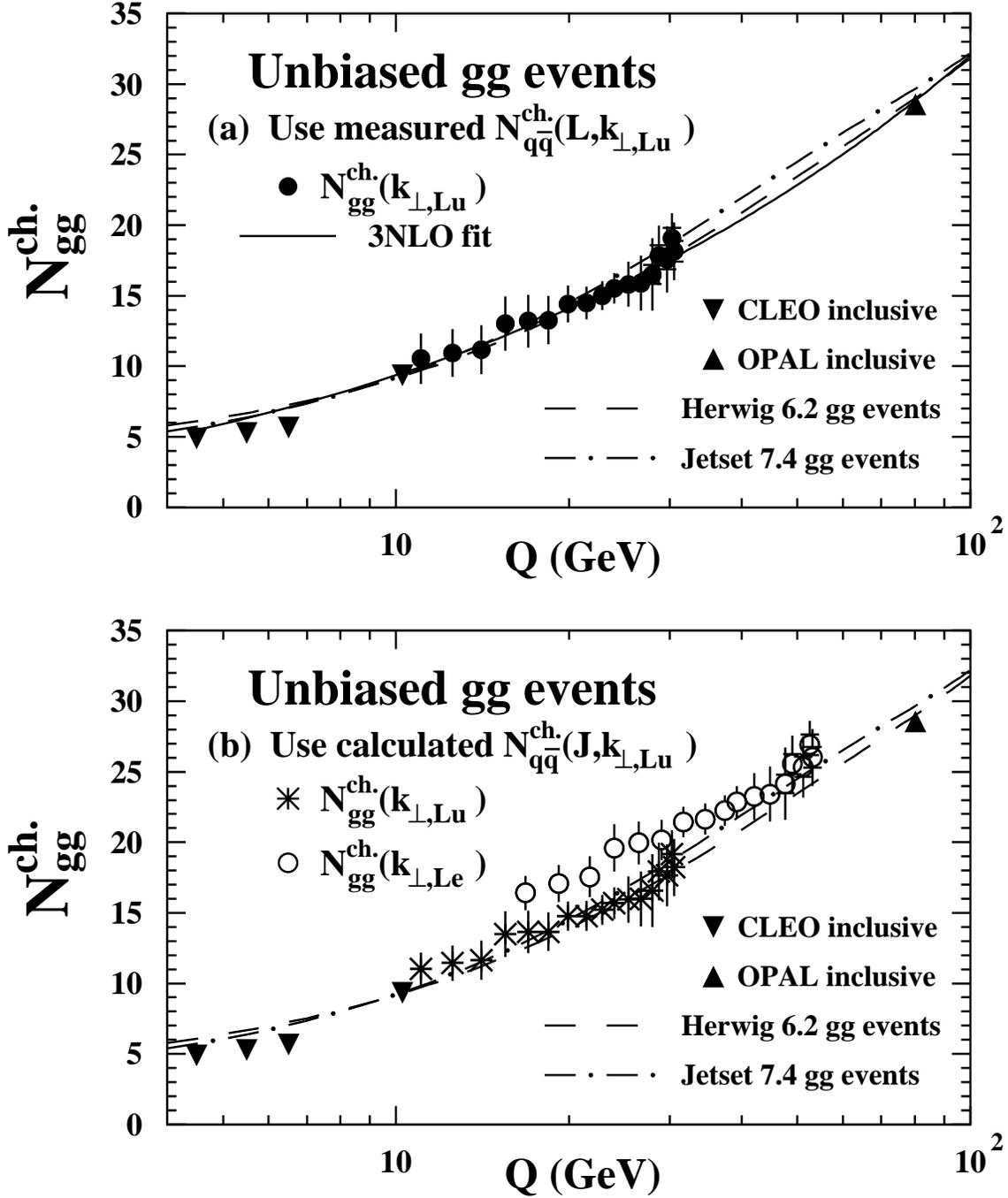}
\end{center}
\vspace*{-.7cm}
\caption{
The mean charged particle multiplicity of unbiased
$gg$ events as a function of scale.
(a)~Results from eq.~(\ref{eq-nggtwo}) using the
measured biased quark jet terms $\nqqch\,(L,\ktlu)$.
The small horizontal bars indicate the statistical uncertainties.
The vertical lines show the total uncertainties,
with statistical and systematic terms added in quadrature.
The solid curve is a fit of the 3NLO expression for
gluon jet multiplicity~\cite{bib-capella,bib-dgaryplb} to the data.
(b)~The corresponding results from eqs.~(\ref{eq-nggtwo})
and~(\ref{eq-nggone}) using the NLO expression for
the biased quark jet terms $\nqqch\,(J,\ktlu)$, 
eq.~(\ref{eq-qbiased}),
where $J$$\,=\,$$L$ for the $\nggch\,(\ktlu)$ results
and $J$$\,=\,$$\lqq$ for the $\nggch\,(\ktle)$ results.
For both (a) and (b),
the triangles show measurements of the inclusive
charged particle multiplicity of unbiased gluon
jets from the CLEO~\cite{bib-cleo92,bib-cleo97} 
and OPAL~\cite{bib-opalgincl98} Collaborations.
The predictions of the Herwig and Jetset Monte Carlo
event generators are also shown.
}
\label{fig-ggmult}
\end{figure}

Fig.~\ref{fig-ggmult} shows our results for the unbiased 
charged particle multiplicities of $gg$ events, $\nggch$.
The solid points in Fig.~\ref{fig-ggmult}a
are obtained from eq.~(\ref{eq-nggtwo})
using the direct measurements of~$\nqqch\,(L,\ktlu)$,
i.e. the solid points in 
Fig.~\ref{fig-nqqbiased}.\footnote{We parametrize 
the data in Fig.~\ref{fig-nqqbiased} using a polynomial.
We then evaluate this polynomial at the scale values
corresponding to our bins in $\theta_1$
(see Table~\ref{tab-scales})
to determine the gluon jet multiplicities
using eq.~(\ref{eq-nggtwo}).}
The small horizontal bars indicate the statistical uncertainties.
The vertical lines show the total uncertainties,
with statistical and systematic terms added in quadrature.
The evaluation of systematic uncertainties is
discussed in Sect.~\ref{sec-systematic}.
The asterisks and open symbols in Fig.~\ref{fig-ggmult}b
show the corresponding results from eqs.~(\ref{eq-nggtwo}) 
and~(\ref{eq-nggone}) using the calculated expressions for
$\nqqch\,(J,\ktlu)$ from eq.~(\ref{eq-qbiased}).
Thus the asterisks in Fig.~\ref{fig-ggmult}b are derived
using the solid curve in Fig.~\ref{fig-nqqbiased},
for example.
The two sets of results from eq.~(\ref{eq-nggtwo})
(solid points in Fig.~\ref{fig-ggmult}a
and asterisks in Fig.~\ref{fig-ggmult}b) 
are seen to be very similar.
This is consistent with the observation in Sect.~\ref{sec-nqqchbiased}
that eq.~(\ref{eq-qbiased}) is generally adequate
for the purposes of our study.

The results from eq.~(\ref{eq-nggtwo}) are plotted
as a function of $Q$$\,=\,$$\ptlu$
and those from eq.~(\ref{eq-nggone}) as a function 
of $Q$$\,=\,$$\ptle$.
The variable $\ptlu$ extends over a smaller range of scale 
than $\ptle$ because the factor of $s$ in the denominator
of the Lund definition of transverse momentum is
always larger in Y events than the factor 
of $\sqq$ in the Leningrad definition,
see eqs.~(\ref{eq-ptlu}) and~(\ref{eq-ptle}).
This effect is also visible in Fig.~\ref{fig-scales}b
and Table~\ref{tab-scales},
i.e.~$\ptlu$ has a maximum of 30.5~GeV compared
to 53.1~GeV for $\ptle$.
Note that with $\theta_1$$\,\geq\,$$35^\circ$ 
as in our standard analysis,
the lower bound on the gluon jet scale is
11.1~GeV for $\ptlu$ and 16.8~GeV for $\ptle$,
see Fig.~\ref{fig-scales}b.

Included in Fig.~\ref{fig-ggmult} are direct measurements 
of the inclusive charged particle multiplicity 
of unbiased gluon jets from the CLEO and OPAL Collaborations,
currently the only results which utilize a hemisphere 
definition of gluon jets.
These results are shown by the triangular symbols.
The three data points at scales of about
4-7~GeV are derived from the hadronic component of
$\Upsilon$(1S)$\,\rightarrow\gamma$$gg$
events~\cite{bib-cleo97}.
The scale $Q$ for these data is given by the 
invariant mass of the hadronic system.
Similarly,
$\Upsilon$(3S)$\,\rightarrow\gamma\chi_{b2}
$(2P)$\,\rightarrow\gamma$$gg$ 
events provide the result at 
$Q$$\,=\,$10.3~GeV~\cite{bib-cleo92},
with the scale given by the $\chi_{b2}$(2P) mass.
The measurement at 
$Q$$\,=\,$80.2~GeV~\cite{bib-opalgincl98}
is determined using hadronic Z$^0$ decays:
Z$^0\rightarrow\,$$q\overline{q}${\gincl},
where {\gincl} refers to a gluon jet hemisphere
recoiling against two almost collinear identified 
quark jets $q$ and $\overline{q}$ in the opposite 
hemisphere~\cite{bib-jwg94}.
For our purposes
the {\gincl} measurement from~\cite{bib-opalgincl98}
has been multiplied by a factor of
two both for the multiplicity and energy scale so
it corresponds to $gg$ ``two-jet events''
analogous to the other data in Fig.~\ref{fig-ggmult}.

Fig.~\ref{fig-ggmult} also shows the predictions of the
Herwig~\cite{bib-herwig59,bib-herwig}
and Jetset Monte Carlo event generators for
the inclusive charged particle multiplicity of $gg$ events
versus $Q$$\,=\,$$\ecm$.
For these results
we use version~6.2~\cite{bib-herwig} of Herwig with 
the following changes to the default parameter set:
the parameter specifying the angular smearing of udsc quark clusters,
CLSMR(1), is set to~0.40,
the parameter controlling the mass spectrum of b quark clusters,
PSPLT(2), is set to~0.33,
and the parameter controlling the production of decuplet baryons,
DECWT, is set to~0.70.
Herwig 6.2 with these changes is found to provide a
good description of the global features of the OPAL data
including those of flavor selected samples.

\begin{table}[tp]
\centering
\begin{tabular}{|c|c|cc|}
 \hline
  & & &  \\[-.2cm]
  $Q$ (GeV) & $\theta_1$ (degrees) & $\nggch\,(\ktlu)$ & $r$  \\[.2cm]
\hline
  & & & \\[-.2cm]
1.23 & 0-5   & $-8.8\pm0.3\pm6.4$& $ -3.1\pm0.1\pm2.1$\\
2.10 & 5-10  & $0.9\pm0.1\pm4.8$&  $ 0.3\pm0.0\pm1.5$\\
3.29 & 10-15 & $4.6\pm0.1\pm2.4$&  $ 1.00\pm0.01\pm0.56$\\
4.88 & 15-20 & $6.9\pm0.1\pm1.8$&  $ 1.25\pm0.01\pm0.34$\\
6.47 & 20-25 & $8.0\pm0.1\pm1.9$&  $ 1.28\pm0.02\pm0.31$\\
8.01 & 25-30 & $8.9\pm0.2\pm1.9$&  $ 1.28\pm0.02\pm0.27$\\
9.52 & 30-35 & $9.7\pm0.2\pm1.9$&  $ 1.29\pm0.02\pm0.26$\\[.2cm]
\hline
  & & & \\[-.2cm]
11.1 & 35-40   & $10.6\pm0.2\pm1.8$   & $ 1.31\pm0.03\pm0.22 $  \\ 
12.6 & 40-45   & $11.0\pm0.3\pm1.7$   & $ 1.29\pm0.03\pm0.20 $  \\ 
14.1 & 45-50   & $11.2\pm0.3\pm1.7$   & $ 1.25\pm0.03\pm0.19 $  \\ 
15.5 & 50-55   & $13.0\pm0.3\pm1.9$   & $ 1.39\pm0.04\pm0.20 $  \\ 
17.0 & 55-60   & $13.2\pm0.4\pm1.8$   & $ 1.35\pm0.04\pm0.19 $  \\ 
18.4 & 60-65   & $13.3\pm0.4\pm1.7$   & $ 1.31\pm0.04\pm0.17 $  \\ 
19.9 & 65-70   & $14.4\pm0.5\pm1.2$   & $ 1.38\pm0.05\pm0.12 $  \\ 
21.4 & 70-75   & $14.5\pm0.5\pm1.1$   & $ 1.344\pm0.046\pm0.098 $  \\ 
22.9 & 75-80   & $15.00\pm0.52\pm0.87$& $ 1.353\pm0.047\pm0.079 $  \\ 
24.0 & 80-85   & $15.53\pm0.57\pm0.94$& $ 1.371\pm0.050\pm0.083 $  \\ 
25.4 & 85-90   & $15.8\pm0.6\pm1.5$   & $ 1.36\pm0.05\pm0.13 $  \\ 
26.7 & 90-95   & $15.9\pm0.6\pm1.9$   & $ 1.34\pm0.05\pm0.16 $  \\ 
27.9 & 95-100  & $16.5\pm0.7\pm2.5$   & $ 1.37\pm0.06\pm0.21 $  \\ 
28.7 & 100-105 & $17.9\pm0.7\pm2.0$   & $ 1.46\pm0.06\pm0.16 $  \\ 
29.6 & 105-110 & $17.6\pm0.7\pm2.3$   & $ 1.42\pm0.06\pm0.18 $  \\ 
30.2 & 110-115 & $19.1\pm0.7\pm1.6$   & $ 1.53\pm0.06\pm0.13 $  \\ 
30.5 & 115-120 & $18.2\pm0.7\pm1.9$   & $ 1.45\pm0.06\pm0.15 $  \\[.2cm]
\hline
\end{tabular}
\caption{
Measurements of unbiased gluon jet multiplicity
$\nggch$ as a function of energy scale $Q$$\,=\,$$\ptlu$,
obtained using eq.~(\ref{eq-nggtwo}).
The corresponding bins of $\theta_1$ in Y events are also indicated.
The results for $\theta_1$$\,\geq\,$35$^\circ$,
listed in the bottom part of the table,
are used in our standard analysis.
These data are shown by the solid points in Fig.~\ref{fig-ggmult}a.
For information,
we also list the results obtained for smaller values of $\theta_1$.
Note that the theoretical formalism used to determine
these results is not expected to be valid for small $\theta_1$.
This explains the negative value of multiplicity derived
for $\theta_1$$\,<\,$$5^\circ$.
The corresponding results for the multiplicity ratio $r$
between gluon and quark jets are also given.
The results for $r$ for $\theta_1$$\,\geq\,$35$^\circ$ are
shown by the solid points in Figs.~\ref{fig-ratior}
and~\ref{fig-allratios}a.
}
\label{tab-nggch}
\end{table} 

\begin{table}[tp]
\centering
\begin{tabular}{|c|c|c|}
 \hline
  & & \\[-.2cm]
  $Q$ (GeV) & $\theta_1$ (degrees) & $\nggch\,(\ktle)$ \\[.2cm]
\hline
  & & \\[-.2cm]
1.59 & 0-5   & $-1.2\pm0.3\pm6.5$ \\
2.90 & 5-10  & $6.8\pm0.1\pm4.8$ \\
4.66 & 10-15 & $9.5\pm0.1\pm2.1$ \\
7.10 & 15-20 & $11.6\pm0.1\pm1.5$ \\
9.58 & 20-25 & $12.9\pm0.1\pm1.5$ \\
12.0 & 25-30 & $14.1\pm0.2\pm1.3$ \\
14.3 & 30-35 & $15.2\pm0.2\pm1.2$ \\[.2cm]
\hline
  & & \\[-.2cm]
16.8 & 35-40 & $16.4\pm0.2\pm1.2$ \\
19.2 & 40-45 & $17.1\pm0.3\pm1.3$ \\
21.8 & 45-50 & $17.6\pm0.3\pm1.4$ \\
24.0 & 50-55 & $19.6\pm0.3\pm1.7$ \\
26.5 & 55-60 & $20.0\pm0.4\pm1.5$ \\
29.0 & 60-65 & $20.2\pm0.4\pm1.4$ \\
31.7 & 65-70 & $21.44\pm0.47\pm0.99$ \\
34.6 & 70-75 & $21.65\pm0.49\pm0.99$ \\
37.4 & 75-80 & $22.25\pm0.52\pm0.95$ \\
39.2 & 80-85 & $22.86\pm0.57\pm0.96$ \\
42.0 & 85-90 & $23.3\pm0.6\pm1.5$ \\
44.8 & 90-95 & $23.4\pm0.6\pm1.8$ \\
47.6 & 95-100 & $24.1\pm0.7\pm2.5$ \\
49.0 & 100-105 & $25.6\pm0.7\pm1.9$ \\
51.1 & 105-110 & $25.4\pm0.7\pm2.1$ \\
52.6 & 110-115 & $26.9\pm0.7\pm1.6$ \\
53.1 & 115-120 & $26.0\pm0.7\pm1.9$ \\[.2cm]
\hline
\end{tabular}
\caption{
Measurements of unbiased gluon jet multiplicity
$\nggch$ as a function of energy scale $Q$$\,=\,$$\ptle$,
obtained using eq.~(\ref{eq-nggone}).
Note that the theoretical formalism used to determine
these results is not expected to be valid for small $\theta_1$.
This explains the negative value of multiplicity derived
for $\theta_1$$\,<\,$$5^\circ$.
The results corresponding to $\theta_1$$\,\geq\,$35$^\circ$
are shown by the open points in Fig.~\ref{fig-ggmult}b.
}
\label{tab-nggchle}
\end{table} 

From Fig.~\ref{fig-ggmult} it is seen that the Monte Carlo
predictions describe the direct measurements of unbiased
gluon jet multiplicity (triangle symbols) well.
The results from the present analysis based on
eq.~(\ref{eq-nggtwo}) 
(solid points in Fig.~\ref{fig-ggmult}a
and asterisks in Fig.~\ref{fig-ggmult}b)
are also well described by the Monte Carlo curves.
In contrast,
the results from eq.~(\ref{eq-nggone}) 
(open symbols in Fig.~\ref{fig-ggmult}b) 
are generally well above the Monte Carlo predictions,
and --~if extrapolated to lower and higher energies~--
appear inconsistent with the direct
measurements from CLEO and OPAL as well.
We note that both Jetset and Herwig have been found to
provide a good description of gluon jet properties,
including multiplicity,
in other studies
(see for example~\cite{bib-opal91}-\cite{bib-delphi96},
\cite{bib-opalgincl98}, \cite{bib-qg95a}).
We therefore conclude that the 
equation based on the Lund definition 
of the gluon jet scale, eq.~(\ref{eq-eden14b}),
yields results which are more consistent with other
studies than the equation based on the 
Leningrad definition, eq.~(\ref{eq-eden14a}).
Our data thus allow a fairly clear discrimination between
the two possibilities eqs.~(\ref{eq-eden14b}) and~(\ref{eq-eden14a}).
We henceforth restrict our analysis of gluon jets to the former
set of results.
Numerical values for the gluon jet measurements
based on eq.~(\ref{eq-eden14b}) (and~(\ref{eq-nggtwo}))
are presented in Table~\ref{tab-nggch}.
The analogous results
based on eq.~(\ref{eq-eden14a}) (and~(\ref{eq-nggone}))
are presented in Table~\ref{tab-nggchle}.
For information,
we include results corresponding to all $\theta_1$ bins
in these tables,
not just the bins above 35$^\circ$ 
used in our standard analysis.
Note that the theoretical formalism is not expected to
be valid for small values of~$\theta_1$.
This explains the negative values of multiplicity determined
for the first bin, 0$\,\leq\,$$\theta_1$$\,\leq\,$5$^\circ$.

\subsection{ The ratios {\boldmath $r$}, {\boldmath $r^{(1)}$}
and {\boldmath $r^{(2)}$} }
\label{sec-ratios}

\subsubsection{Determination of the ratios}
\label{sec-determination}

Having established that the formalism of~\cite{bib-eden} 
yields consistent results with the direct measurements
of unbiased gluon jet multiplicity 
if the equation for three-jet event multiplicity
based on the Lund definition of transverse momentum
is employed (Sect.~\ref{sec-gluonmult}),
we proceed to a comparison of the unbiased gluon and quark 
jet terms and to a test of the theoretical 
expressions for the ratios $r$, $r^{(1)}$ and $r^{(2)}$,
see eqs.~(\ref{eq-one})-(\ref{eq-three}).
For this purpose we use the gluon jet data
shown by the solid points in Fig.~\ref{fig-ggmult}a
since these results are derived using the direct measurements
of $\nqqch\,(L,\ktlu)$ rather than the NLO expression, 
eq.~(\ref{eq-qbiased}).
They therefore rely on fewer theoretical assumptions.

We employ the following procedure to determine
$r$, $r^{(1)}$ and~$r^{(2)}$ from experiment.
The corrected unbiased quark and gluon jet multiplicities,
shown by the solid points in Figs.~\ref{fig-nchepem}
and~\ref{fig-ggmult}a,
are separately fitted using the 3NLO expressions for 
$\nq$ and $\ng$~\cite{bib-capella,bib-dgaryplb},
respectively.
To better constrain the results,
the measurements of inclusive gluon jet multiplicity
from CLEO at 10.3~GeV and OPAL at 80.2~GeV
are included in the gluon jet fit.
The fitted parameters are $\Lambda_{eff.}$ 
and the normalization constant~$K$.
(For the quark jet data,
this is the same fit discussed in Sect.~\ref{sec-procedure}.)
The results of the fits are shown by the solid curves
in Figs.~\ref{fig-nchepem} and~\ref{fig-ggmult}a.
The fitted parameters and corresponding
$\chi^2$ per degree-of-freedom
are listed in Table~\ref{tab-fitpar}.
The $\chi^2$ per d.o.f. and overall description of the data
are seen to be reasonable in both cases.
The ratio of the fitted expressions for 
$\ng$ and $\nq$ defines~$r$, eq.~(\ref{eq-one}).
We calculate the first and second derivatives of
the analytic equations for $\ng$ and $\nq$ with respect to $y$
and evaluate the resulting expressions using the corresponding
parameter values from Table~\ref{tab-fitpar}.
The ratios of these terms define~$r^{(1)}$ and~$r^{(2)}$,
eqs.~(\ref{eq-two}) and~(\ref{eq-three}).

\begin{figure}[t]
\begin{center}
  \begin{tabular}{c}
    \epsfxsize=15cm
    \epsffile{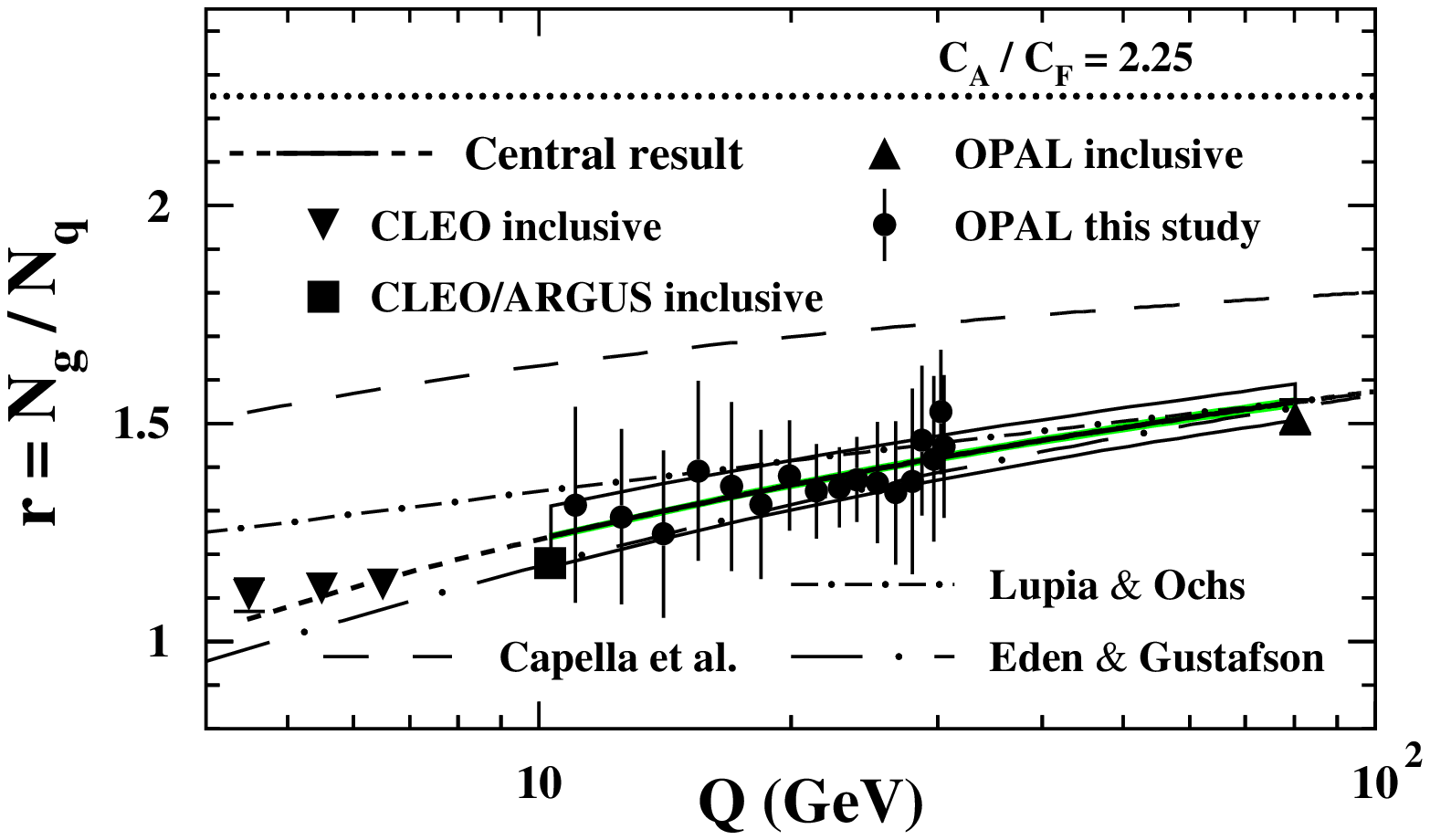} \\
  \end{tabular}
\end{center}
\caption{
Results for the ratio $r$ of the mean charged 
particle multiplicities between unbiased gluon and 
uds quark jets as a function of scale.
The central result is shown by the solid curve;
the shaded and open bands show the corresponding 
statistical and total uncertainties, respectively
(the statistical uncertainties are barely visible).
The short-dashed curve is an extrapolation of the fit.
The solid points show measurements based on gluon
jet results from the present analysis.
The triangle and square symbols show results
based on direct measurements of multiplicity in unbiased jets
from the CLEO~\cite{bib-cleo92,bib-cleo97},
ARGUS~\cite{bib-argus92} and OPAL~\cite{bib-opalgincl98}
Collaborations.
The dash-dotted and long-dashed curves show theoretical predictions
based on analytic~\cite{bib-capella,bib-eden} 
and numerical~\cite{bib-lupia} techniques.
}
\label{fig-ratior}
\end{figure}

\begin{figure}[tp]
\begin{center}
  \begin{tabular}{c}
    \epsfxsize=15cm
    \epsffile{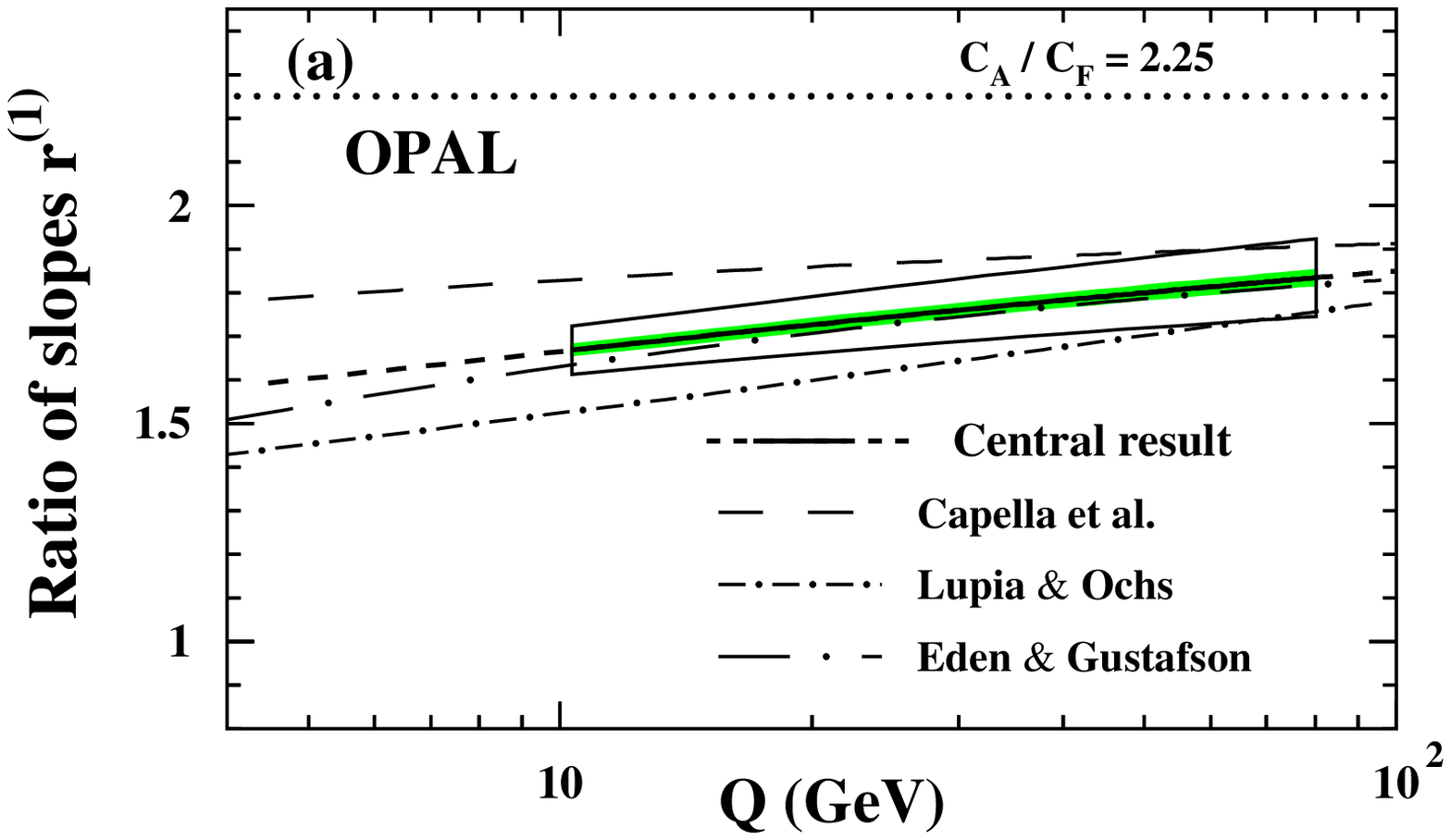} \\
    \epsfxsize=15cm
    \epsffile{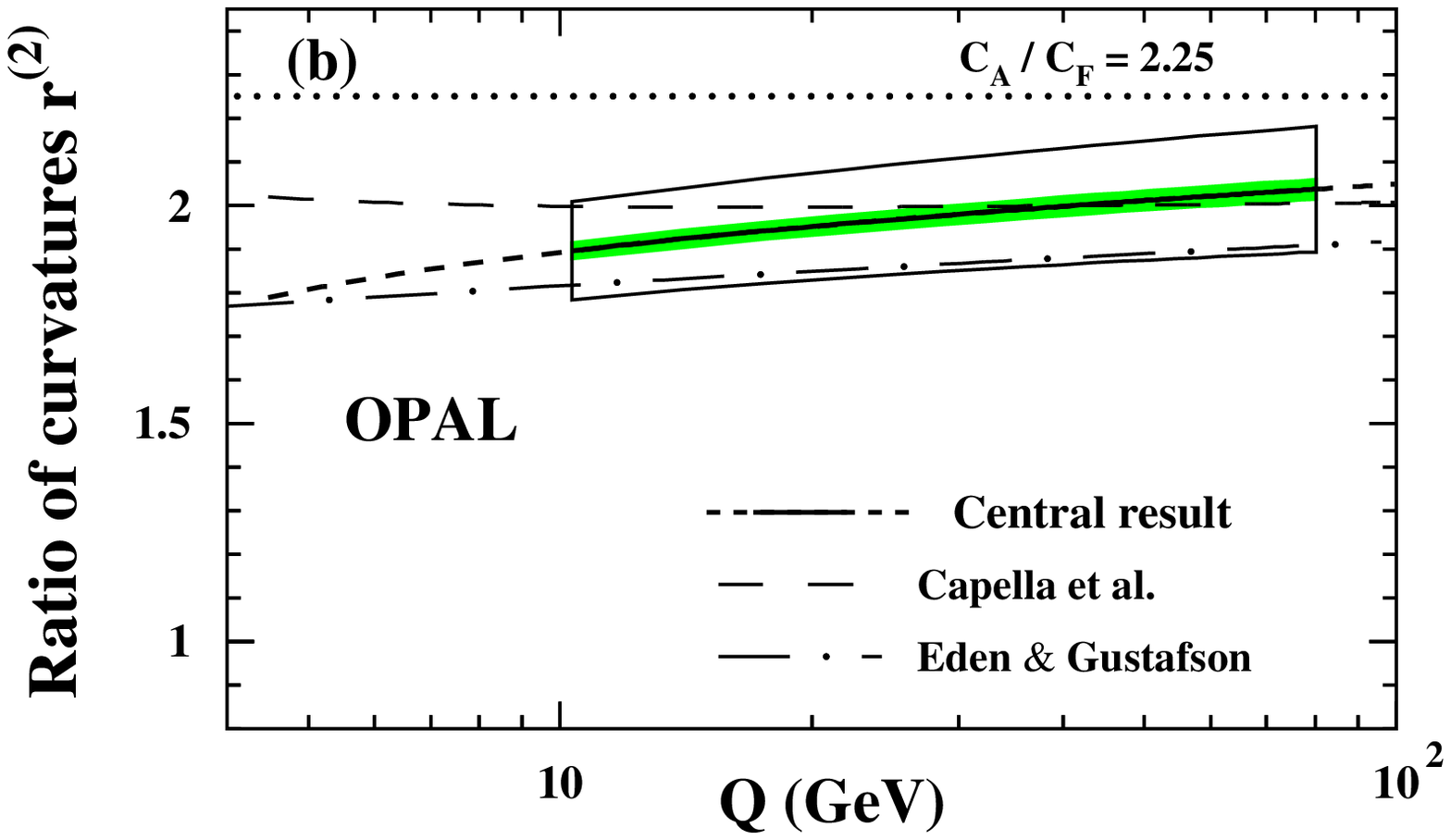} \\
  \end{tabular}
\end{center}
\caption{
(a) and~(b):
The ratios of slopes $r^{(1)}$ and of curvatures~$r^{(2)}$
between unbiased gluon and uds quark jets 
as a function of scale.
The curves and bands have the
same meaning as in Fig.~\ref{fig-ratior}.
}
\label{fig-r1r2}
\end{figure}

Our results for $r$ are shown in Fig.~\ref{fig-ratior},
and those for $r^{(1)}$ and~$r^{(2)}$ in Fig.~\ref{fig-r1r2}.
The central results are indicated by solid curves.
The shaded bands show the statistical uncertainties.
The overall uncertainties,
with statistical and systematic terms added in quadrature,
are shown by the open bands.
The bands cover the energy range of data in the 
gluon jet fit,
from 10.3 to 80.2~GeV.
The short-dashed curves show extrapolations 
of the results to lower and higher scales.
At 30~GeV,
a typical scale for the gluon jet data in these figures,
the ratios are determined to be:
\begin{eqnarray}
%  r\,(10.3\,{\mathrm GeV})       & = & 1.240\pm 0.005 \pm 0.zz 
%      \label{eq-r10} \\
%  r^{(1)}\,(10.3\,{\mathrm GeV}) & = & 1.667\pm 0.011 \pm 0.zz 
%      \label{eq-r110} \\
%  r^{(2)}\,(10.3\,{\mathrm GeV}) & = & 1.895\pm 0.018 \pm 0.zz  
%      \label{eq-r210}
  r\,(30\,{\mathrm GeV})       & = & 1.422\pm 0.006 \pm 0.051 
     \;\;\;\; ,      \label{eq-r10} \\
  r^{(1)}\,(30\,{\mathrm GeV}) & = & 1.761\pm 0.013 \pm 0.070 
     \;\;\;\; ,      \label{eq-r110} \\
  r^{(2)}\,(30\,{\mathrm GeV}) & = & 1.98\pm 0.02 \pm 0.13  
      \label{eq-r210}
   \;\;\;\; ,
\end{eqnarray}
where the first uncertainty is statistical and
the second systematic.
At 80~GeV,
corresponding to the highest energy gluon jet data in the fits,
the results are:
\begin{eqnarray}
  r\,(80\,{\mathrm GeV})       & = & 1.548\pm 0.008 \pm 0.041 
     \;\;\;\; ,      \label{eq-r30} \\
  r^{(1)}\,(80\,{\mathrm GeV}) & = & 1.834\pm 0.016 \pm 0.088 
     \;\;\;\; ,      \label{eq-r130} \\
  r^{(2)}\,(80\,{\mathrm GeV}) & = & 2.04\pm 0.02 \pm 0.14  
      \label{eq-r230}
   \;\;\;\; .
\end{eqnarray}
Analogous results for $r$ versus scale were 
previously published in~\cite{bib-delphi96,bib-delphi99,bib-opal00}
and for $r^{(1)}$ in~\cite{bib-delphi99,bib-opal00}.
These earlier studies were based on biased jet samples, however.
Therefore the quantitative results from those studies
cannot be compared to theoretical predictions without ambiguity.
Our result for $r^{(2)}$ is the first
experimental determination of that quantity.

Included in Fig.~\ref{fig-ratior} are the results
for $r$ derived from the individual gluon jet
measurements in Fig.~\ref{fig-ggmult}a.
The solid points in Fig.~\ref{fig-ratior} are
obtained by dividing the results shown by the solid
points in Fig.~\ref{fig-ggmult}a by the results
of the 3NLO parametrization of quark jet multiplicity at
the corresponding scales.
Numerical values of these results for $r$ 
are compiled in Table~\ref{tab-nggch}.
The triangle and square symbols in Fig.~\ref{fig-ratior}
are obtained as follows:
(i)~for the OPAL data point at 80.2~GeV,
the published result~\cite{bib-opalgincl98} is used,
based on uds quark events analogous
to the present study;
(ii)~for the three CLEO points at 4-7~GeV,
the published results~\cite{bib-cleo97} are used
after applying a correction obtained from Jetset
to remove the contributions of c events from
the quark jet samples;
(iii)~for the point labelled ``CLEO/ARGUS,''
the CLEO measurement of
gluon jet multiplicity at 10.3 GeV~\cite{bib-cleo92}
is divided by the ARGUS measurement of $\nchee$ at 
10.5~GeV~\cite{bib-argus92}
(see Fig.~\ref{fig-nchepem}),
after applying a correction to account for c and b events 
in the quark jet sample
and for the small energy difference between 
the gluon and quark jet results.

As a consistency check,
we also determined the ratios $r$ and $r^{(1)}$
using the technique employed in~\cite{bib-opal00}.
The corrected quark jet results
(solid points in Fig.~\ref{fig-nchepem})
were fitted using a polynomial expression:
\begin{equation}
  \nqq = a_0 + a_1\,L + a_2\,L^2
     \;\;\;\; ,
  \label{eq-nqqmadjid}
\end{equation}
with $L$ given in eq.~(\ref{eq-lll}),
and where $a_0$, $a_1$ and~$a_2$ are fitted parameters.
We find 
$a_0$$\,=\,$$13.20\pm 0.67$,
%$a_1$$\,=\,$$-2.913\pm 0.119$ and
$a_1$$\,=\,$$-2.91\pm 0.12$ and
$a_2$$\,=\,$$0.2818\pm 0.0057$,
where the uncertainties are statistical.
The $\chi^2$ (d.o.f.) of the fit is $3.74$~(15).
We then fitted the 19 gluon jet measurements
in our analysis
(solid points in Fig.~\ref{fig-ggmult}a
and the 10.3 and 80.2~GeV results from
CLEO and OPAL)
using the expression:
\begin{equation}
  \ngg = b_0 + b_1\,\nqq
     \;\;\;\; ,
  \label{eq-nggmadjid}
\end{equation}
with the results 
$b_0$$\,=\,$$-4.07\pm 0.27$,
%$b_0$$\,=\,$$-4.074\pm 0.270$,
$b_1$$\,=\,$$1.759\pm 0.030$
and $\chi^2$ (d.o.f.)$\,=\,$$31.5$~(17).
The result for~$r$,
given by the ratio of eqs.~(\ref{eq-nggmadjid}) 
and~(\ref{eq-nqqmadjid}),
is indistinguishable from the
solid curve in Fig.~\ref{fig-ratior}.
The result for $r^{(1)}$,
obtained by differentiation of eq.~(\ref{eq-nggmadjid}),
is, by construction, a constant:
$r^{(1)}$$\,\equiv\,$$b_1$$\,=\,$$1.759\pm 0.030$~(stat.).
This result is consistent with that shown
in Fig.~\ref{fig-r1r2}a,
see also eq.~(\ref{eq-r110}).
Not only is $r^{(1)}$ constant with this method,
but $r^{(2)}$$\,\equiv\,$$b_1$$\,=\,$$r^{(1)}$,
i.e. no distinction is made between
the ratios of slopes and curvatures.
Therefore this alternate technique does not provide 
as much information as our standard procedure
described above.

To obtain a consistency check of our
result for~$r^{(2)}$,
we applied the following procedure.
The 19 gluon jet measurements
%--~the 17 solid points in Fig.~\ref{fig-ggmult}a
%along with the CLEO and OPAL results at
%10.3 and 80.2~GeV~--
were clustered into 16 groups of 
four contiguous measurements.
%with each group defined by contiguous data points in
%scale~$Q$.
Thus,
labelling the data points from 1 to 19 in order of
increasing~$Q$,
the first group is composed of data points 1-4,
the second of 2-5, etc.
A straight line fit was performed for each group:
\begin{equation}
  (\ngg)_i = (a_g)_i\,L + (b_g)_i
   \;\;\;\;\;\; i=1,16  \;\;\;\; ,
  \label{eq-nggline}
\end{equation}
with $a_g$ and $b_g$ the fitted parameters.
By fitting to four points at a time,
rather than e.g. to two points only,
the influence of bin-to-bin fluctuations is reduced.
We then evaluated the fitted polynomial expression
eq.~(\ref{eq-nqqmadjid})
at the scale values of the gluon jet measurements
to obtain 19 corresponding results for quark jets,
which were then fitted in the same manner:
\begin{equation}
  (\nqq)_i = (a_q)_i\,L + (b_q)_i
   \;\;\;\;\;\; i=1,16  \;\;\;\; .
  \label{eq-nqqline}
\end{equation}
A scale value was associated with each slope point
$(a_g)_i$ or $(a_q)_i$ using the arithmetic mean of the four $Q$
values in the fits.
%eqs.~(\ref{eq-nggline}) and~(\ref{eq-nqqline}).
As a systematic check,
we also defined these scales
using the weighted means of the $Q$ values of the four
associated gluon jet measurements.
This alternate method yielded consistent results
with those presented below.
The slopes $(a_g)_i$ and $(a_q)_i$ were in turn fitted
using straight lines:
\begin{equation}
  a_g = c_g\,L + d_g
  \label{eq-ggslope}
\end{equation}
\begin{equation}
  a_q = c_q\,L + d_q
  \label{eq-qqslope}
\end{equation}
with $c_i$ and $d_i$ ($i$$\,=\,$$q,g$)
the fitted terms.
The ratio $r^{(2)}$ found using this technique is a constant
given by $r^{(2)}$$\,=\,$$c_g/c_q$.
We find $r^{(2)}$$\,=\,$$2.25\pm0.16\,$(stat.),
consistent with our result in Fig.~\ref{fig-r1r2}b,
see also eq.~(\ref{eq-r210}).

The procedure described in the previous paragraph can
also be used to obtain a second consistency check for~$r^{(1)}$.
Taking the weighted mean of the 16 results for the
ratio of slopes $r^{(1)}_i$$\,=\,$$(a_g/a_q)_i$
yields
$r^{(1)}$$\,=\,$$1.68\pm0.02\,$(stat.),
consistent with the result in eq.~(\ref{eq-r110})
to within about 1.0 standard deviation of
the total uncertainties.

We note that the method to determine $r^{(1)}$ and $r^{(2)}$
based on 
eqs.~(\ref{eq-nggline})-(\ref{eq-qqslope})
is more direct than our standard procedure because
no functional form is assumed for how the gluon jet multiplicity
varies with scale.
The resulting uncertainties are considerably larger,
however,
and no information on the scale dependence of
$r^{(1)}$ and $r^{(2)}$ is obtained.
Therefore we retain the analysis based on the
3NLO parametrizations
of multiplicities for our standard results.

Analytic expressions for $r$, $r^{(1)}$ and~$r^{(2)}$,
valid to 3NLO,
were recently presented by Capella et al.~\cite{bib-capella}.
All three ratios are predicted to converge to the
ratio of QCD color factors, $\caa$/$\cff$$\,=\,$2.25,
in the asymptotic limit of large jet energies.
This convergence is predicted to be more rapid for $r^{(1)}$
than $r$~\cite{bib-capella,bib-delphi99}
and yet more rapid for $r^{(2)}$~\cite{bib-capella}.
Our data are in agreement with this prediction,
i.e.~$r$$\;<\;$$r^{(1)}$$\,<\,$$r^{(2)}$$\,<\,$2.25
for the scales accessible in our study,
see eqs.~(\ref{eq-r10})-(\ref{eq-r230}).
Note that we also observe the hierarchy
$r$$\;<\;$$r^{(1)}$$\,<\,$$r^{(2)}$
using the alternate analysis strategies summarized
by eqs.~(\ref{eq-nqqmadjid})-(\ref{eq-qqslope}).

\begin{figure}[tp]
\begin{center}
  \begin{tabular}{c}
    \epsfxsize=15cm
    \epsffile{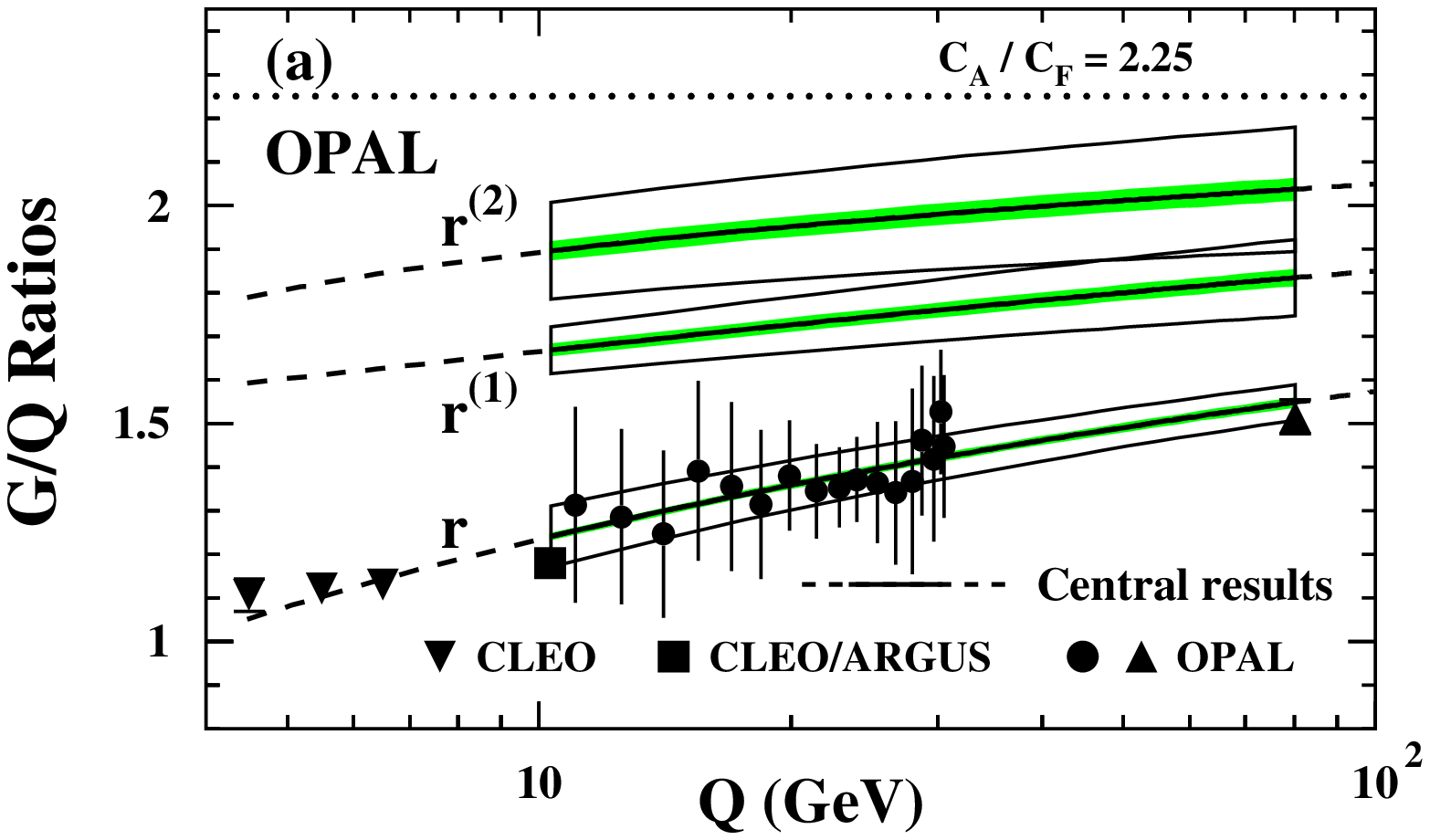} \\
    \epsfxsize=15cm
    \epsffile{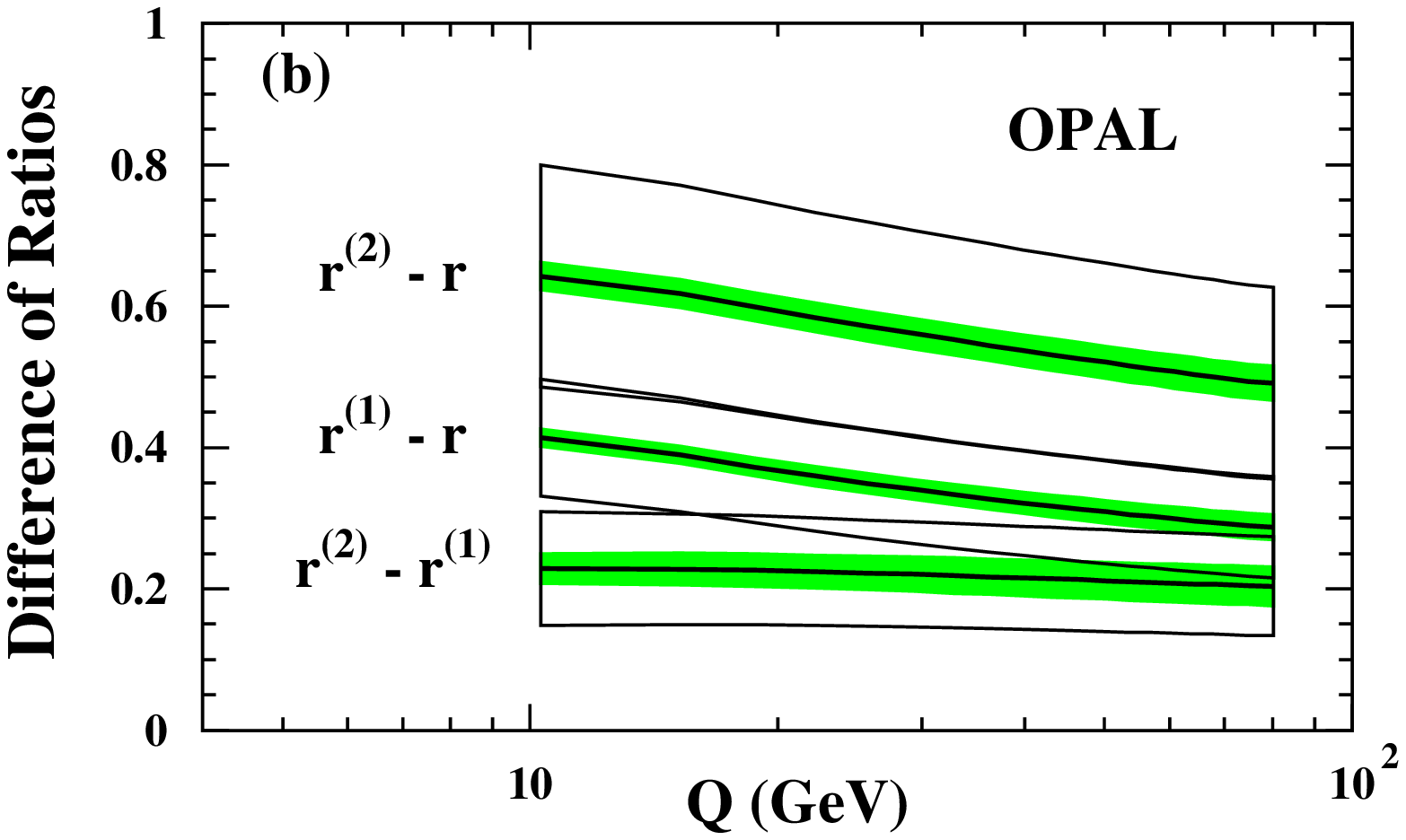} \\
  \end{tabular}
\end{center}
\caption{
(a) The experimental results for 
$r$, $r^{(1)}$ and $r^{(2)}$ from 
Figs.~\ref{fig-ratior} and~\ref{fig-r1r2},
collected together;
(b)~the differences between ratios
$r^{(2)}-r$, $r^{(1)}-r$ and $r^{(2)}-r^{(1)}$
as a function of scale.
The central results are shown by solid curves;
the shaded and open bands show the corresponding 
statistical and total uncertainties, respectively.
}
\label{fig-allratios} 
\end{figure}

To make the comparison of our results for 
$r$, $r^{(1)}$ and~$r^{(2)}$ more clear,
the data from Figs.~\ref{fig-ratior} and~\ref{fig-r1r2}
are displayed together in Fig.~\ref{fig-allratios}a.
The dominant uncertainties in our analysis are systematic
(see Sect.~\ref{sec-systematic})
and are correlated between the three ratios.
Therefore,
in Fig.~\ref{fig-allratios}b,
we present results for the differences
$r^{(2)}$$\,-\,$$r$, $r^{(1)}$$\,-\,$$r$ and $r^{(2)}$$\,-\,$$r^{(1)}$,
for which correlated systematic uncertainties partially cancel.
At 30~GeV, we find:
\begin{eqnarray}
  \left[\, r^{(2)}-r \,\right]\,(30\,{\mathrm GeV}) 
         & = & 0.56\pm 0.02 \pm 0.14
     \;\;\;\; , \label{eq-r2r} \\
  \left[\, r^{(1)}-r \,\right]\,(30\,{\mathrm GeV}) 
         & = & 0.341\pm 0.014 \pm 0.075
     \;\;\;\; , \label{eq-r1r} \\
  \left[\, r^{(2)}-r^{(1)} \,\right]\,(30\,{\mathrm GeV}) 
         & = & 0.221\pm 0.024 \pm 0.070
     \;\;\;\; , \label{eq-r2r1}
\end{eqnarray}
where the first uncertainty is statistical and
the second systematic.
These results demonstrate that $r^{(2)}$ exceeds $r^{(1)}$ by
3.0 standard deviations of the total uncertainties,
with yet more significant deviations from
zero observed for the other two differences of ratios.
%eqs.~(\ref{eq-r2r}) and~(\ref{eq-r1r}).

\subsubsection{Comparison to theoretical expressions}

The 3NLO analytic
predictions of Capella et al.~\cite{bib-capella} 
for $r$, $r^{(1)}$ and $r^{(2)}$
are shown by the long-dashed curves in 
Figs.~\ref{fig-ratior} and~\ref{fig-r1r2}.
For $Q$$\,=\,$30~GeV,
the ratios are predicted 
to be 1.73, 1.87 and 2.00, respectively,
assuming $\nff$$\,=\,$5 and $\Lambda$$\,=\,$0.20~GeV.
Thus, the analytic predictions for $r$ and $r^{(1)}$
exceed the corresponding
experimental results in eqs.~(\ref{eq-r10})-(\ref{eq-r110})
by about 22\% and~6\%,
while the theory agrees with the data,
eq.~(\ref{eq-r210}), for~$r^{(2)}$.
Therefore the analytic predictions agree better with the data the 
higher the level of differentiation of the multiplicity curves.
This suggests that higher order corrections
are smaller for $r^{(2)}$ than $r^{(1)}$,
and for $r^{(1)}$ than~$r$.

We note there are ambiguities in the analytic calculation
related to the effective number of active quark flavors and the
scale $Q$ at which to evaluate the strong coupling strength.
The 3NLO predictions shown here are based on $n_f$$\,=\,$5
(as stated above) and $Q$$\,=\,$$\ecm$.
Other plausible choices such as $n_f$$\,=\,$4 and $Q$$\,=\,$$\ecm$/4
change the analytic prediction for $r$ by as much as about 10\%
and can bring the theoretical curve into better agreement with 
the data,
see Fig.~2 of~\cite{bib-opalgincl96} for example.

A second QCD prediction for $r$ versus scale was
recently presented by Lupia and Ochs~\cite{bib-lupia}.
This prediction is based on numerical,
rather than analytic, techniques,
incorporating more accurate phase space limits for the emission
of soft gluons and a more complete treatment of energy
conservation than the analytic expression.
Unlike the analytic result,
a prediction is presented only for $r$, however,
not for $r^{(1)}$ or~$r^{(2)}$.
We derive a prediction for $r^{(1)}$ from the numerical 
calculation in the following manner.
We begin with numerical predictions for the individual
gluon and quark jet multiplicities,\footnote{We thank
Wolfgang Ochs for providing these results.}
given in steps of 0.02 units in~$y$
with $y$$\,=\,$$\ln\, (Q/Q_c)$ where 
$Q_c$$\,=\,$$0.507$~GeV~\cite{bib-lupia}.
We make linear interpolations between the points 
and use the slopes of these lines to estimate
${\mathrm{d}}\ng /{\mathrm{d}}y$ and
${\mathrm{d}}\nq /{\mathrm{d}}y$,
which are then used to determine~$r^{(1)}$ as
a function of scale~$Q$.
We attempted to extend this technique to obtain
a prediction for~$r^{(2)}$ as well.
The derived results for $r^{(2)}$ were
very scattered from point to point as a function of~$Q$
because of limited numerical precision.
We therefore do not quote a prediction for $r^{(2)}$
based on the calculation of~\cite{bib-lupia}.

%$\ng$ and $\nq$ are
%treated in the manner described above for the data,
%i.e.~they are separately fitted using the corresponding
%3NLO expressions to find effective values of $\Lambda_{eff.}$
%and~$K$,
%then the ratio of the expressions and ratios
%of their first and second derivatives are evaluated
%using the fitted parameter values
%to define $r$, $r^{(1)}$ and~$r^{(2)}$.

The predictions of the numerical calculation for $r$ 
and $r^{(1)}$ are shown by the short-dashed-dotted curves in
Figs.~\ref{fig-ratior} and~\ref{fig-r1r2}a.
At 30~GeV,
the numerical calculation predicts $r$ to be~1.45,
in agreement with our measurement at that scale,
eq.~(\ref{eq-r10}).
The corresponding result for $r^{(1)}$ is 1.64,
about 7\% below the data,
eq.~(\ref{eq-r110}).
For $r$,
the agreement with data is therefore better for 
the numerical method than for the analytic calculation.
The level of agreement of the two predictions with data
is similar for $r^{(1)}$.
The difference 
%of about 15\% 
between the analytic
and numerical results for $r$
(long-dashed and short-dash-dotted curves in Fig.~\ref{fig-ratior})
is believed to arise from the differences in phase space 
limits and energy conservation noted above.
For further discussion, see~\cite{bib-dgphysrep}.

Finally,
theoretical predictions for $r$, $r^{(1)}$ and~$r^{(2)}$
can be derived from the formalism of
Ed\'{e}n and Gustafson~\cite{bib-eden},
using eq.~(\ref{eq-gluonevolution}).
To obtain a prediction for $r^{(1)}$,
we begin with the experimental results for the
slope of quark jets,
${\mathrm{d}}\nqq(L) /{\mathrm{d}}L$,
determined in the manner described in Sect.~\ref{sec-determination}
for ${\mathrm{d}}\nqq(y) /{\mathrm{d}}y$,
with $y$$\,=\,$$L/2$.
We insert this result in eq.~(\ref{eq-gluonevolution})
to obtain a prediction for 
${\mathrm{d}}\ngg(L) /{\mathrm{d}}L$.
Dividing this result by the measured
${\mathrm{d}}\nqq(L) /{\mathrm{d}}L$
yields the prediction for~$r^{(1)}$.
To obtain a prediction for $r^{(2)}$,
we differentiate eq.~(\ref{eq-gluonevolution}) with respect
to $L$ and follow an analogous procedure to that described
for~$r^{(1)}$.
% in the previous three sentences.
To obtain a prediction for $r$,
we fix the gluon jet multiplicity at 80.2~GeV to
the {\gincl} measurement from OPAL~\cite{bib-opalgincl98}.
We then use the prediction for 
${\mathrm{d}}\ngg(L) /{\mathrm{d}}L$ determined as
described above to evaluate the gluon jet multiplicity 
at other scales and divide the result by the
3NLO expression for the quark jet multiplicity
fitted to data
(solid curve in Fig.~\ref{fig-nchepem}).

The predictions obtained in this manner
are shown by the long-dash-dotted curves in 
Figs.~\ref{fig-ratior} and~\ref{fig-r1r2}.
The results are seen to be in good 
overall agreement with the data.
We note, however,
that these predictions are based on the
experimental measurements of quark jet multiplicities.
Furthermore,
the prediction for $r$ utilizes the measured gluon jet
multiplicity at $Q$$\,=\,$80.2~GeV as stated above.
%Furthermore,
%the predictions are based on the same formalism~\cite{bib-eden}
%used to extract the measurements of gluon jet multiplicity
%(Sect.~\ref{sec-gluonmult}).
Therefore,
the predictions we derive based on~\cite{bib-eden}
are not independent of the data.

\subsection{Measurement of {\boldmath $\caa/\cff$} }
\label{sec-cacf}

Measurements of the energy dependence of
unbiased particle multiplicities in gluon and quark jets 
can also be used to determine an effective value of
the ratio of QCD color factors $\caa/\cff$.
Such a possibility was first discussed
in~\cite{bib-delphi99}.
To perform this measurement,
we integrate eq.~(\ref{eq-gluonevolution})
to obtain:
\begin{equation}
  \ngg\,(L) = \ngg\,(L_0) + \frac{\caa}{\cff}\,{\cal F}\,(L)
   \;\;\;\; ,
  \label{eq-nggintegrate}
\end{equation}
with $\ngg\,(L_0)$ an integration constant which
serves as the reference point for $\ngg\,(L)$,
and where
\begin{equation}
  {\cal F}\,(L) \equiv {\displaystyle \int^{L-c_g+c_q}_{L_0-c_g+c_q} }
    \left[ \left( 1-\frac{\alpha_0 c_r}{x} \right)
       \frac{ {\mathrm d}\,\nqq\,(x) }{ {\mathrm d}x } 
            \right] {\mathrm d}x
   \;\;\;\; .
  \label{eq-fintegrate}
\end{equation}
We evaluate eq.~(\ref{eq-fintegrate}) numerically 
using the experimental results
for ${\mathrm d}\nqq\,(Q)/{\mathrm d}Q$
discussed in Sect.~\ref{sec-determination}.
As the reference point $\ngg\,(L_0)$ in eq.~(\ref{eq-nggintegrate}),
we use the direct measurement of gluon jet multiplicity
from OPAL~\cite{bib-opalgincl98}.
Thus $L_0$$\,=\,$$2\ln\,(Q_0/\Lambda)$,
with $Q_0$$\,=\,$80.2~GeV
and $\ngg\,(L_0)$$\,=\,$28.56~\cite{bib-opalgincl98}.
For $\Lambda$ we used the fitted result for
$\Lambda_{eff.}$
for quark jets from Table~\ref{tab-fitpar},
$\Lambda$$\,=\,$0.19~GeV.
Note that the results for ${\caa}/{\cff}$ are independent
of this choice.

A one parameter fit of eq.~(\ref{eq-nggintegrate})
is made to the gluon jet measurements
in Fig.~\ref{fig-ggmult}a,
excluding the CLEO data below 7~GeV,
to obtain
\begin{equation}
  \frac{\caa}{\cff} = 2.232\pm 0.008\,({\mathrm stat.})
     \pm 0.108\,({\mathrm syst.})
     \pm 0.082\,({\mathrm ref.})
   \;\;\;\; ,
  \label{eq-cacf}
% total uncertainty 0.136
\end{equation}
where the first uncertainty is statistical.
The second uncertainty is a systematic term
evaluated as described in Sect.~\ref{sec-systematic}.
The third uncertainty is an additional systematic
term defined by the difference observed if
the CLEO measurement of $\ngg$ at 10.3~GeV
is used as the reference point,
rather than the OPAL measurement at 80.2~GeV.
%Note that the choice of the normalization point
%represents a significant uncertainty to the final result.
The $\chi^2$ (d.o.f.) of the fit is 29.5 (17).
The result of our fit, eq.~(\ref{eq-cacf}),
is consistent with the QCD value
$\caa/\cff$$\,=\,$2.25.

\begin{figure}[t]
\begin{center}
  \epsfxsize=14cm
  \epsffile{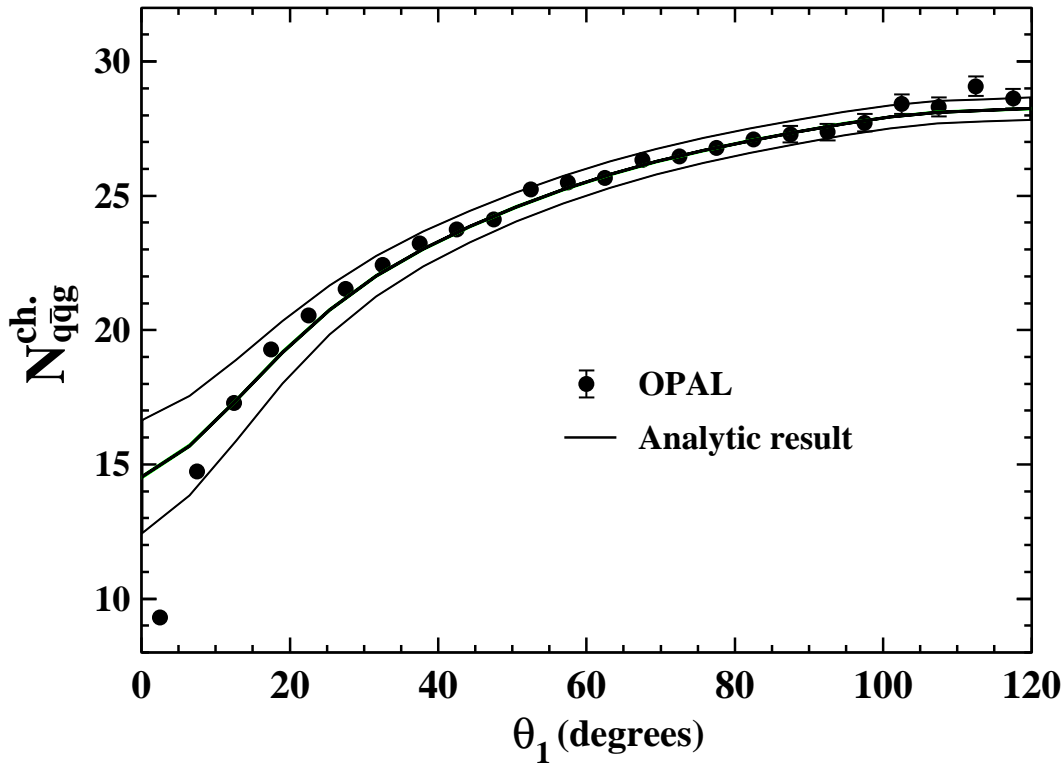}
\end{center}
\caption{
The mean charged particle multiplicity
of three-jet uds flavor Y events from Z$^0$ decays
compared to the analytic expression,
eq.~(\ref{eq-eden14b}),
for which the gluon jet multiplicity term $\ngg\,(\ktlu)$ is
determined from a one parameter 
fit of eq.~(\ref{eq-nggintegrate})
to unbiased gluon jet data and the quark jet term
$\nqq\,(L,\ktlu)$ is measured.
The open band shows the evaluated
uncertainty of the analytic result.
Note the theory is not expected to be applicable for small
values of $\theta_1$ because the events in this case do not
have a well defined three jet event structure.
}
\label{fig-qqgfitted}
\end{figure}

The fitted expression of eq.~(\ref{eq-nggintegrate}) 
yields a curve which is almost identical to the 3NLO curve
in Fig.~\ref{fig-ggmult}a:
therefore we do not display the fitted curve of
eq.~(\ref{eq-nggintegrate}) in addition.
In Fig.~\ref{fig-qqgfitted} we present the
result for the event multiplicity in Y events.
The data in this figure are the same as
in Fig.~\ref{fig-scales}a.
The curve is obtained from eq.~(\ref{eq-eden14b}),
with $\ngg\,(\ktlu)$ the fitted expression of
eq.~(\ref{eq-nggintegrate})
and $\nqq\,(L,\ktlu)$ the measured result
from Sect.~\ref{sec-nqqchbiased}.
The open band shows the uncertainty of the analytic result,
evaluated using the uncertainty from eq.~(\ref{eq-cacf}) 
in addition to those shown for the $\nqq\,(L,\ktlu)$ terms
in Fig.~\ref{fig-nqqbiased}.
The curve is seen to provide an accurate 
description of the data to within its uncertainty
for $\theta_1$ values larger than approximately 10$^\circ$.
This represents a considerable improvement compared to
the results of~\cite{bib-jwgdelphi} based on earlier
theoretical formalism~\cite{bib-dok88},
in which the corresponding fitted analytic expression
was found to describe the data for
$\theta_1$ values larger than about 60$^\circ$ only
(see Fig.~4 of~\cite{bib-jwgdelphi}).

To test our technique of determining $\caa/\cff$,
we apply our analysis to parton level Monte Carlo events 
%at both the parton and hadron levels,
generated with a c.m. energy of 10~TeV.
The parton level is defined using final-state partons,
i.e. those which are present after the termination
of the parton shower.
We use Herwig version 6.2 with the parameters
discussed in Sect.~\ref{sec-gluonmult}.
Applying our analysis to parton level events with 
a very large energy tests our method to determine $\caa/\cff$ 
since the parton level asymptotic result must be consistent 
with 2.25 for the analysis to be considered valid,
see~\cite{bib-jwgdelphi} for further discussion.
%This test is necessary to ensure that the similarity
%of our result eq.~(\ref{eq-cacf}) with the QCD value
%of $\caa/\cff$ is not merely coincidental.
%Herwig generally predicts that QCD variables in 
%e$^+$e$^-$ annihilations reach
%their asymptotic values at c.m. energies of
%several TeV or more.
%At the parton level,
The parton level Herwig events at 10~TeV yield
$\caa/\cff$$\,=\,$$2.285\pm 0.010\,({\mathrm stat.})\pm 0.047\,({\mathrm ref.})$,
%$\caa/\cff$$\,=\,$$2.285\pm 0.010^{+0}_{-0.086}\,({\mathrm ref.})$,
where the central value is defined using the Herwig parton level
prediction for $\ngg$ at 80~GeV as the reference point.
The first uncertainty is statistical
while the second uncertainty is defined 
by the difference found if the Herwig
prediction for $\ngg$ at 10~GeV 
is used as the reference point instead
(i.e.~this is analogous to how the third uncertainty 
of eq.~(\ref{eq-cacf}) is defined).
%The second uncertainty is defined by the maximum signed 
%difference if the Herwig prediction at 10~GeV or 10~TeV 
%is used as the reference point instead of the prediction at 80~GeV.
%The corresponding result at the hadron level is 
%$\caa/\cff$$\,=\,$$2.285\pm0.009\pm 0.006\,({\mathrm ref.})$.
%$\caa/\cff$$\,=\,$$2.29\pm0.01^{+0}_{-0.15}\,({\mathrm ref.})$.
Thus our one parameter fit method yields an asymptotic result
of 2.25 to within the uncertainties associated with the
analysis procedure,
implying that the similarity of our measurement
eq.~(\ref{eq-cacf}) with the QCD value of $\caa/\cff$
is not merely coincidental.
%This result is consistent with 2.25,
%implying that our one parameter fit method 
%represents a theoretically consistent technique
%to determine~$\caa/\cff$.
For information,
the parton level Herwig result at $\ecm$$\,=\,$91~GeV
is $2.03\pm0.01\,({\mathrm stat.})$ if the 
prediction for $\ngg$ at 80~GeV is used as the reference point
and $2.35\pm0.02\,({\mathrm stat.})$
if the prediction at 10~GeV is used.
The analogous results for hadron level Herwig events are 
$2.34\pm0.01$ and $2.07\pm0.02$, respectively.
We therefore observe a much stronger dependence on 
the choice of the normalization point for the 91~GeV
Monte Carlo results than for the 10~TeV results
or for the data.
The Herwig results at 91~GeV are nonetheless 
consistent with 2.25,
once the systematic uncertainty related to choice of the
normalization point is considered.

%Following the suggestions in~\cite{bib-delphi99},
%we also fitted eq.~(\ref{eq-nggintegrate})
%to the gluon jet data after subtracting/adding
%a constant offset term $N_0$,
%with $\caa/\cff$ and $N_0$ as the fitted parameters.
%We obtained
%$\caa/\cff$$\,=\,$$z.zzz\pm z.zzz \pm z.zzz \pm z.zzz$
%and 
%$N_0$$\,=\,$$z.zzz\pm z.zzz \pm z.zzz \pm z.zzz$
%where the uncertainties have the same meaning as
%for eq.~(\ref{eq-cacf}).
%The $\chi^2$ (d.o.f.) of the fit is zz.z (17).
%The result for $\caa/\cff$ is again very similar
%to the QCD value 2.25.
%However when we apply this alternative method to
%Herwig events at 10~TeV,
%we obtain
%$\caa/\cff$$\,=\,$$2.285^{+0}_{-0.086}\,({\mathrm ref.})$,
%at the parton level and
%$\caa/\cff$$\,=\,$$2.285^{+0}_{-0.086}\,({\mathrm ref.})$,
%at the hadron level.
%Since the two parameter fit method does not yield the
%correct asymptotic result of 2.25,
%%%%we do not consider it  a valid technique
%it is not clear it represents a valid technique
%to determine~$\caa/\cff$.

%The hadron level result is consistent with our measurement,
%eq.~(\ref{eq-cacf}),
%to within the uncertainties.
%The prediction for the hadronization correction,
%given by the ratio of the parton to hadron level results,
%is found to depend strongly on the choice of the normalization point,
%being 1.14, 0.86 or 1.00 if the normalization point is chosen at 
%10~GeV, 80~GeV or 10~TeV, respectively.
%Therefore we prefer not to apply a hadronization
%correction to our result, eq.~(\ref{eq-cacf}).

\section{Systematic uncertainties}
\label{sec-systematic}

To evaluate systematic uncertainties,
the analysis was repeated with the
following changes to the standard analysis.
The full differences between the standard results
and those found using each of these changes
were used to define symmetric systematic uncertainties.
For items~1, 5 and~7, listed below,
the largest of the described differences
with respect to the standard result
was assigned as the systematic uncertainty.
The uncertainties were added in quadrature
to define the total systematic uncertainties.

For the measurement of the gluon jet multiplicity
$\nggch$ (Fig.~\ref{fig-ggmult}),
the following changes were made.
Items 1-3 and 5 in this list were also applied
to the study of two-jet events,
see Sect.~\ref{sec-nqqchbiased}.
The items are listed in roughly decreasing order
of the size of their contribution to the total 
uncertainties of the $\nggch$ measurements.
\begin{enumerate}
\item  The jet finding was performed using the
  Cambridge and Luclus jet finders,
  rather than the Durham jet finder.
  For the Luclus jet finder,
  we did not perform a direct measurement
  of the biased quark jet terms $\nqqch\,(J,\ktlu)$
  since it is not obvious how to relate the resolution
  scale $d_{join}$ to $\ptlu$
  as mentioned in Sect.~\ref{sec-nqqchbiased}.
  Instead we used the analytic expression, eq.~(\ref{eq-qbiased}),
  for the determination of the gluon jet multiplicities $\nggch\,(L)$.
  The difference between the Durham and Cambridge results
  was used to define
  the systematic uncertainty due to jet finding for the
  two jet event measurements shown in Fig.~\ref{fig-nqqbiased}.
\item  Tracks selected for the uds tagging procedure
  were required to have a signed impact parameter
  which satisfied $\dsign/\sigma_{\dsign}$$\,>\,$1.5,
  rather than $\dsign/\sigma_{\dsign}$$\,>\,$3.0;
  at the same time we required
  $|\cos (\theta_{\mathrm{thrust}})|$ to be less than~0.8,
  rather than 0.9.
  The purpose of these two changes was to increase the
  purity of tagged uds events;
  the estimated uds event purity increased to 85.1\% from 78.5\%
  while the number of selected events decreased to $8610$ 
  from $22\,365$.
  The more restrictive condition on $|\cos (\theta_{\mathrm{thrust}})|$
  yields a higher uds purity because a larger
  fraction of tracks in an event traverses
  the silicon microvertex detector.
\item  The Herwig Monte Carlo event generator
  was used to determine the corrections
  for initial-state photon radiation, detector response,
  and uds event and gluon jet misidentification,
  rather than Jetset.
  For this purpose,
  we used version 5.9 of the program~\cite{bib-herwig59}
  with the parameter values in~\cite{bib-qg95b},
  except that the cluster mass
  cutoff CLMAX was increased from 3.40 to 3.75~GeV/$c^2$ to
  improve the model's description of the mean charged
  particle multiplicity in inclusive hadronic Z$^0$ decays.
\item For our measurements of the gluon jet multiplicities
  $\nggch$ based on eq.~(\ref{eq-nggtwo}),
  we used the analytic expression, eq.~(\ref{eq-qbiased}),
  rather than the direct measurement of $\nqqch\,(L,\ktlu)$.
\item  Charged tracks alone were used for the data and
  Monte Carlo samples with detector simulation,
  rather than charged tracks plus electromagnetic clusters.
  As an additional check on the track selection,
  the standard analysis (tracks and clusters) was varied by
  increasing the minimum transverse momentum of charged tracks
  with respect to the beam axis
  from from 0.05~GeV/$c$ to 0.15~GeV/$c$.
\end{enumerate}
For the two-jet events (Fig.~\ref{fig-nqqbiased}),
the dominant systematic term was from item~1 at low $\ptlu$
and from item~3 at intermediate and high $\ptlu$.

In addition to the above,
the following changes were made to evaluate systematic
uncertainties for the ratios $r$, $r^{(1)}$ and $r^{(2)}$.
\begin{enumerate}
\item[6.] The ARGUS and 91~GeV LEP measurements of $\nchee$
  were excluded from the fit of quark jet multiplicity
  (solid curve in Fig.~\ref{fig-nchepem}).  
  These two data points have the smallest uncertainties
  of the results in the fit and
  were excluded to potentially maximize a systematic variation.
  The results of the alternate fit are
  $\Lambda_{eff.}$$\,=\,$$0.261\pm 0.061$,
  $K$$\,=\,$$0.154\pm 0.014$
  and $\chi^2$ (d.o.f.)$\,=\,$~1.5~(14).
  Besides the evaluation of a systematic uncertainty for
  $r$, $r^{(1)}$ and $r^{(2)}$,
  this alternate result for $\nqqch\,(L)$ was used to derive
  the NLO prediction shown by the dashed curve in Fig.~\ref{fig-nqqbiased}.
\item[7.] 
  Y events with 
  $50^{\circ}$$\,\leq\,$$\theta_1$$\,\leq\,$$120^{\circ}$
  were used in the fit of gluon jet multiplicity versus scale,
  rather than
  $35^{\circ}$$\,\leq\,$$\theta_1$$\,\leq\,$$120^{\circ}$;
  as an alternate check,
  the ratios were determined after increasing the inclusive gluon jet
  measurement from CLEO at 10.3~GeV by its
  one standard deviation total uncertainty,
  then decreasing it by one standard deviation,
  and repeating the same operation for the inclusive gluon
  jet measurement from OPAL at 80.2~GeV.
\end{enumerate}
For the ratios,
the largest systematic terms were from items~1, 6 and 7.
The contributions of items~2 and~5 were
by comparison negligible,
with the other items intermediate.
For our measurement of $\caa/\cff$ (Sect.~\ref{sec-cacf}),
increasing and decreasing the inclusive OPAL measurement at 
80.2~GeV by one standard deviation (item~7) 
provided the largest contribution
to the overall uncertainty.
Item~1 provided the second largest contribution while
the contributions of the other items were much smaller.

For the data listed in Tables~\ref{tab-nggch} and~\ref{tab-nggchle}
and the corresponding results in Figs.~\ref{fig-ggmult},
\ref{fig-ratior} and~\ref{fig-allratios}a,
the systematic uncertainty evaluated for each bin
was averaged with the results from its two neighbors
to reduce the effect of bin-to-bin fluctuations.
The single neighbor was used for bins on the endpoints
of the distributions.

\section{Summary}

We have presented measurements of the mean charged particle
multiplicity of three-jet ``Y events''
as a function of the opening angle between the 
two lowest energy jets, $\theta_1$,
and of two-jet events as a function of the resolution
scale separating the two- and three-jet event classes.
The measurements were performed using the Durham jet finder.
The Cambridge and Luclus jet finders
were used as systematic checks for the three-jet measurements
and the Cambridge jet finder for the two-jet measurements.
The two- and three-jet events were selected from
a sample of Z$^0$ decays to light flavor (u, d or s)
quark-antiquark pairs produced in {\epem} annihilations.
The restriction to light quark flavors improves
the correspondence of the data to theoretical predictions.
%for particle multiplicities in jets.

We use these data to test recent theoretical 
formalism~\cite{bib-eden,bib-edenkhoze}
for particle multiplicity in jets which accounts for biases 
introduced by the jet finding criteria.
We find that the theoretical prediction for particle 
multiplicity in biased two-jet events,
eq.~(\ref{eq-qbiased}),
agrees fairly well with our two-jet event data.
In conjunction with the results from Y events,
the two-jet data are used to extract the multiplicity of
{\it unbiased} gluon jets at a variety of scales.
Of two possible forms proposed in~\cite{bib-edenkhoze} to
obtain this information
--~one~\cite{bib-eden} based on the definition of 
gluon jet transverse momentum from the Lund group and the 
other~\cite{bib-dok88} from the Leningrad group~--
we find clear preference for the former because of the consistency 
of the derived results with direct measurements 
of unbiased gluon jet multiplicity 
from the CLEO~\cite{bib-cleo92,bib-cleo97} 
and OPAL~\cite{bib-opalgincl96}-\cite{bib-opalgincl98} Collaborations
as well as with Monte Carlo expectations.

The unbiased gluon jet multiplicities extracted with this technique,
which range in scale from 11.1 to 30.5~GeV,
are compared to corresponding results from light 
unbiased quark jets.
In conjunction with direct measurements of unbiased
gluon jet multiplicity at 10.3 and 80.2~GeV from CLEO and OPAL,
we determine the ratio of charged particle multiplicities between
gluon and quark jets, $r$$\,=\,$$\ng/\nq$,
and the corresponding ratios of slopes and of curvatures,
$r^{(1)}$$\,=\,$$({\mathrm{d}}\ng /{\mathrm{d}}y)
/({\mathrm{d}}\nq / {\mathrm{d}}y)$
and
$r^{(2)}$$\,=\,$$({\mathrm{d}}^2\ng /{\mathrm{d}}y^2)
/({\mathrm{d}}^2\nq / {\mathrm{d}}y^2)$,
as a function of energy scale
$y$$\,=\,$$\ln\,(Q/\Lambda)$,
with $Q$ the jet energy and $\Lambda$ the QCD scale parameter,
set to 0.20~GeV in our study.
At 30~GeV,
a typical scale in our analysis,
we find
$r$$\,=\,$$1.422\pm0.006\pm0.051$,
$r^{(1)}$$\,=\,$$1.761\pm0.013\pm0.070$ and
$r^{(2)}$$\,=\,$$1.98\pm0.02\pm0.13$,
and at 80~GeV
$r$$\,=\,$$1.548\pm0.008\pm0.041$,
$r^{(1)}$$\,=\,$$1.834\pm0.016\pm0.088$ and
$r^{(2)}$$\,=\,$$2.04\pm0.02\pm0.14$,
where the first uncertainty is statistical and
the second systematic.
These results are based on QCD analytic 
parametrizations~\cite{bib-capella,bib-dgaryplb}
of the scale dependence of particle multiplicity in jets.
We also present results which utilize polynomial parametrizations,
or which assume no functional dependence for the growth
of gluon jet multiplicity with scale,
and obtain consistent results albeit with larger uncertainties.
Our results are
in agreement with the QCD prediction that $r^{(2)}$ should
be closer to its asymptotic value of 2.25 than $r^{(1)}$,
and $r^{(1)}$ closer to 2.25 than~$r$,
for the finite energies of our experiment.
Our result for $r^{(2)}$ is the first experimental determination
of that quantity.

The results for $r$, $r^{(1)}$ and $r^{(2)}$
are compared to recent QCD predictions
based on analytic~\cite{bib-capella}
and numerical~\cite{bib-lupia} techniques,
as well as to predictions derived from
the formalism of~\cite{bib-eden}.
We find overall agreement between the
experimental and theoretical results.
%Given that no corrections are applied to account for
%hadronization,
%and that the theoretical results employ perturbative
%evolution down to scales around typical hadron masses
%($\sim$0.5~GeV),
%this provides evidence for the applicability of
%perturbative methods down to low energy scales
%and for the hypothesis of Local Parton Hadron Duality,
%see for example~\cite{bib-lphd} for a discussion of
%these last issues.

Finally,
we use the results on the energy dependence of
$\ng$ and $\nq$ to determine
an effective value of the ratio of QCD color factors,
$\caa/\cff$.
We find
$\caa/\cff$$\,=\,$$2.23\pm0.14\,$(total),
consistent with the QCD value of 2.25.

\section{Acknowledgments}

We thank Igor Dremin, Patrick Ed\'{e}n
and Valery Khoze for helpful comments.

We particularly wish to thank the SL Division for the efficient operation
of the LEP accelerator at all energies
 and for their close cooperation with
our experimental group.  We thank our colleagues from CEA, DAPNIA/SPP,
CE-Saclay for their efforts over the years on the time-of-flight and trigger
systems which we continue to use.  In addition to the support staff at our own
institutions we are pleased to acknowledge the\\[2mm]
Department of Energy, USA, \\
National Science Foundation, USA, \\
Particle Physics and Astronomy Research Council, UK, \\
Natural Sciences and Engineering Research Council, Canada, \\
Israel Science Foundation, administered by the Israel
Academy of Science and Humanities, \\
Minerva Gesellschaft, \\
Benoziyo Center for High Energy Physics,\\
Japanese Ministry of Education, Science and Culture (the
Monbusho) and a grant under the Monbusho International
Science Research Program,\\
Japanese Society for the Promotion of Science (JSPS),\\
German Israeli Bi-national Science Foundation (GIF), \\
Bundesministerium f\"ur Bildung und Forschung, Germany, \\
National Research Council of Canada, \\
Research Corporation, USA,\\
Hungarian Foundation for Scientific Research, OTKA T-029328, 
T023793 and OTKA F-023259,\\
Fund for Scientific Research, Flanders, F.W.O.-Vlaanderen, Belgium.\\

\end{document}